\documentclass[fleqn,usenatbib]{mnras}

\usepackage{newtxtext,newtxmath}

\usepackage[T1]{fontenc}
\usepackage{subcaption}
\usepackage{booktabs}
\usepackage[table,xcdraw]{xcolor}
\usepackage{array}  
\usepackage{tabularx}

\usepackage{xcolor}
\usepackage[normalem]{ulem}

\definecolor{RefRed}{HTML}{C00000}
\definecolor{RefTeal}{HTML}{008C95}

\DeclareRobustCommand{\VAN}[3]{#2}
\let\VANthebibliography\thebibliography
\def\thebibliography{\DeclareRobustCommand{\VAN}[3]{##3}\VANthebibliography}

\usepackage{graphicx}	
\usepackage{amsmath}	
\usepackage{todonotes}
\usepackage{tikz}
\usetikzlibrary{arrows.meta,positioning,fit,calc}

\newcommand{\Msun}{\ensuremath{\mathrm{M}_\odot}}
\newcommand{\Fable}{\textsc{Fable}}
\newcommand{\vect}[1]{\boldsymbol{#1}}
\defcitealias{buttigieg_2025}{SB25}

\title[The GWB from \Fable{}]{Comparing gravitational wave background predictions from cosmological simulations to pulsar timing observations}

\author[S. Buttigieg et al.]{
Stephanie Buttigieg,$^{1,2}$\thanks{E-mail: sb2583@cam.ac.uk (SB)}
Debora Sijacki,$^{1,2}$,
Christopher J.\ Moore $^{1,2,3}$, Martin A. Bourne $^{2,4}$, \newauthor Alberto Sesana $^{5,6,7}$
\\
$^{1}$Institute of Astronomy, University of Cambridge, Madingley Road, Cambridge CB3 0HA, UK\\
$^{2}$Kavli Institute for Cosmology, Cambridge (KICC), University of Cambridge, Madingley Road, Cambridge CB3 0HA, UK\\
$^{3}$Department of Applied Mathematics and Theoretical Physics, Centre for Mathematical Sciences, University of Cambridge, \\ Wilberforce Road, Cambridge, CB3 0WA, UK\\
$^{4}$Centre for Astrophysics Research, Department of Physics, Astronomy and Mathematics, University of Hertfordshire, \\ College Lane, Hatfield, AL10 9AB, UK \\
$^{5}$ Dipartimento di Fisica ``G. Occhialini'', Universit\`a di Milano - Bicocca, Piazza della Scienza 3, 20126 Milano, Italy \\
$^{6}$ INAF, Osservatorio Astronomico di Brera, via Brera 20, Milano, 20121, Italy \\
$^{7}$ INFN, Sezione di Milano-Bicocca, Piazza della Scienza 3, 20126 Milano, Italy
}

\date{MNRAS, accepted}

\pubyear{\the\year{}}

\begin{document}
\label{firstpage}
\pagerange{\pageref{firstpage}--\pageref{lastpage}}
\maketitle

\begin{abstract}
The recent detection of a gravitational wave background (GWB) by pulsar timing arrays (PTAs) may represent the first evidence of gravitational waves from merging supermassive black hole binaries, opening a new window on the low-frequency end of the gravitational wave spectrum. 
These inspiralling binaries are expected to dominate the signal, although most theoretical models seem to predict somewhat lower amplitudes than what is observed. 
We present the first comprehensive statistical framework to quantify the tension between PTA measurements and theoretical predictions, maximising the constraining power of current data and allowing straightforward application to future PTA datasets.
We further investigate how different assumptions in the observational inference, particularly the use of a power-law model for the GWB spectrum, can bias tension estimates and potentially overstate discrepancies with theory. 
We apply our framework to compare predictions from the \Fable{} cosmological simulation with the NANOGrav 15-year dataset. 
For our fiducial black hole population, we find tension values of $1\sigma$–$2.5\sigma$, indicating no statistically significant disagreement with the observations. 
We further explore physically motivated modifications to the merging black hole population, guided by electromagnetic observations and theoretical uncertainties. 
In particular, scenarios with boosted black hole masses at high redshift and more equal-mass mergers substantially increase the predicted GWB amplitude, improving agreement with PTA data. 
Finally, we investigate the high-mass end of the black hole mass function and the impact of finite simulation volume. We find that the $(100 \, \mathrm{cMpc} \, h^{-1})^3$ \Fable{} box is sufficient to robustly predict the GWB signal at the most constraining frequency.
\end{abstract}




\section{Introduction} \label{sec:introduction}

Pulsar timing arrays (PTAs) have begun probing the nanohertz (nHz) regime of the gravitational wave (GW) spectrum by monitoring correlated timing residuals from arrays of millisecond pulsars. The measured angular dependence, consistent with the quadrupolar form predicted by the Hellings–Downs (HD) curve, likely constitutes evidence for a stochastic GW background (GWB)~\citep{hellings_1983}. Various different collaborations, including the European, Indian, Chinese and Parkes PTAs~\citep{antoniadis_2023,Reardon_2023,Xu_2023,Rana_2025}, together with the North American Nanohertz Observatory for Gravitational Waves (NANOGrav)~\citep{Agazie_2023b}, have detected this signal at levels of significance between $2\sigma$ and $4\sigma$. Their independent analyses are consistent with each other, with discrepancies being lower than $1\sigma$~\citep{agazie_2024}.

One of the most likely sources of the GWB is a cosmological population of inspiralling supermassive black hole binaries (SMBHBs)~\citep{rajagopal_1995, jaffe_2003, wyithe_2003, sesana_2008, ravi_2015}. Supermassive black holes (SMBHs), with masses ranging between $10^5$~\Msun{} and $10^{11}$~\Msun{}, are found at the centre of most massive galaxies~\citep{magorrian_1998, richstone_1998, Kormendy_2013}, and observational evidence suggests that they co-evolve with their host galaxies, likely via a variety of feedback processes~\citep{magorrian_1998, gebhardt_2000, Ferrarese_2002, Termaine_2002, Marconi_2003, haring_2004, gultekin_2009, Kormendy_2013, McConnell_2013}. Within the hierarchical $\Lambda$CDM model of structure formation~\citep{white_1978}, galaxies are expected to merge with each other~\citep{Toomre_1972, barnes_1992, boylan_2005, springel_2005b, rodriguez_2015}, with surveys confirming that such mergers are common~\citep{lin_2004, ryan_2008, duncan_2019}. Consequently, SMBH binaries are expected to form, and growing evidence from electromagnetic (EM) observations now points to the existence of dual and binary SMBHs~\citep{Rodriguez_2006,koss_2012,Comerford_2013,mannucci_2023,Ubler_2024, perna_2025}. After various dynamical hardening processes that extract energy from the binary and shrink the separation, the SMBH binary could reach separations where GW radiation becomes significant and drives the two SMBHs towards coalescence~\citep{begelman_1980}. A superposition of signals generated by binaries at various stages of their evolution would then source the GWB detected by PTAs~\citep{Kelley_2016,Kelley_2017,agazie_2023}. 

The predicted GWB from a population of SMBHBs is shaped by a wide range of interconnected factors. These include the SMBH mass function, which in turn depends on scaling relations between host galaxies and their central black holes~(BHs), the distribution of binary mass ratios, the hardening mechanisms and associated timescales that govern merger rates, the evolution of binary eccentricities, and the redshift distribution of merger events. In recent years, significant theoretical effort has been devoted to modelling SMBHB populations using a variety of approaches. Semi-analytic galaxy formation models combine empirically or theoretically motivated prescriptions for galaxy and SMBH growth with merger trees derived from cosmological structure formation, allowing detailed predictions for SMBHB merger rates and properties~\citep{somerville_2008,sesana_2009,izquierdo_2022}. Cosmological hydrodynamical simulations evolve galaxies and SMBHs self-consistently within a cosmological volume, capturing the complex interplay of gas dynamics, star formation, feedback, and SMBH physics. SMBHB populations can then be extracted from these simulations and combined with post-processing prescriptions for sub-resolution hardening~\citep{Kelley_2016, volonteri_2020, buttigieg_2025, chen_2025}. Other models aim to incorporate BH dynamics directly on-the-fly in simulations such as RAMCOAL, integrated in the RAMSES code ~\citep{Li_2024, Li_2026}, and KETJU~\citep{rantala_2017, Mannerkoski_2023} that has been applied in cosmological zoom-in simulations~\citep{mannerkoski_2021, Mannerkoski_2022, keitaanranta_2026}.  Phenomenological parameterisations take a more flexible approach, describing SMBHB populations using simplified, tunable functions for merger rates, mass distributions, and eccentricities, calibrated to observations or simulations; these models allow rapid exploration of parameter space~\citep{sesana_2013b, chen_2017, chen_2019,Sato-Polito_2024, Sato-Polito_2025}.

These methods depend on, and are often calibrated against, the BH mass function (BHMF), which at present is constrained almost entirely by EM observations. In the local Universe ($z \lesssim 0.1$), spatially resolved measurements of stellar and gas kinematics around SMBHs enable direct mass determinations~\citep{shapiro_2006,walsh_2012,davis_2013,onishi_2015}. These are complemented by EM surveys that detect low-luminosity active galactic nuclei (AGN), tracing SMBHs in low-accretion states~\citep{ho_1997, Ho_1999,Terashima_2003, nagar_2005}. Combining these detections with the well-established correlations between SMBH mass and properties of the host galaxy allows the BHMF to be inferred from the broader galaxy population~\citep{marconi_2004,graham_2007,greene_2007,Kormendy_2013, McConnell_2013}. At high redshift, limited angular resolution makes direct mass measurements incredibly challenging, and our understanding of the SMBH population relies largely on the So\l tan-type argument, which links the integrated AGN luminosity density to the total mass accreted over cosmic time~\citep{soltan_1982}. However, mass estimates derived from AGN luminosities remain sensitive to assumptions about radiative efficiency and the fraction of SMBHs that are actively accreting~\citep{yu_2002, davis_2011}. Interestingly, recent observations from the \textit{James Webb Space Telescope (JWST)} are now providing a new insight into this high-$z$ population through detections of broad emission lines and virial BH mass estimates calibrated on low-redshift sources. These early \textit{JWST} results hint at a possible evolution in the SMBH–host galaxy scaling relations~\citep{Pacucci_2023, Bogdan_2024, juodzbalis_2024, Maiolino_2024, jones_2025}, whereby at high redshifts SMBHs are more massive at a given host galaxy mass than their counterparts in the local Universe. 

Based on current estimates of the BHMF, the measured spectral slope of the GWB is consistent with a signal originating from a population of SMBHBs, following the expected scaling of $h_c(f) \propto f^{-2/3}$~\citep{phinney_2001}. However, many models developed prior to the PTA detections predicted amplitudes somewhat lower than those now inferred from the observations, as discussed in more detail in Section~\ref{sec:tension} and shown in Fig.~\ref{fig:model_comparison}~\citep{agazie_2023, chen_2025, matt_2025}. 

Previous studies have started quantifying the tension between PTA observations and theoretical predictions to statistically assess this apparent underprediction. For example, \citet{Sato-Polito_2024} reported a $2$–$4.5\sigma$ discrepancy at the `reference' frequency of $1\,\mathrm{yr}^{-1}$ when comparing PTA observations with predictions based on the local SMBH population inferred from scaling relations. They attributed this tension to an insufficient number of the most massive SMBHs, suggesting that roughly an order of magnitude more high-mass BHs would be required to reproduce the observed signal, although the presence of even a few ultra-massive systems is disfavoured, as such binaries would likely be individually resolvable~\citep{Sato-Polito_2025}.

In this work, we revisit the tension between the \Fable{} predictions and the NANOGrav 15-year data in greater detail. Our goals are twofold: (i) to carefully quantify the level of discrepancy between simulation and observation, and (ii) to explore how different physical properties of the SMBHB population shape the GWB. In particular, we investigate the impact of hardening timescales and mass ratios on the signal. We also examine the influence of the BHMF from two complementary perspectives: (a) a mass–relation approach, in which we vary BH masses at fixed host galaxy properties, and (b) a convergence test to assess how rare, extremely massive SMBHs, that are potentially missing from a limited cosmological volume, could contribute to the observed background. 
We will show that, while the fiducial \Fable{} model does predict a slightly lower amplitude than what is observed, the resulting tension is not statistically significant. Furthermore, physically motivated modifications and observationally supported scenarios offer substantial flexibility, allowing the predicted signal to rise and reach full consistency with the measured GWB amplitude.

The remainder of this paper is organised as follows. In Section~\ref{sec:methods}, we describe the \Fable{} simulation, focusing on its catalogue of SMBH mergers, and outline the use of the \texttt{holodeck} package~\citep{Kelley_2017} to generate the corresponding GWB signal. In Section~\ref{sec:tension}, we examine different approaches for quantifying the tension between theoretical predictions and current observations, and evaluate the tension between the \Fable{} predictions and the NANOGrav 15-year dataset using the method detailed in Appendix~\ref{appendix:difference_vector}. Section~\ref{sec:astro_uncert} discusses key astrophysical uncertainties affecting the theoretical modelling of the SMBH population and SMBHB hardening processes, and how these influence the predicted GWB. In Section~\ref{sec:volume}, we assess the impact of finite simulation volume on the predicted GWB and test whether the \Fable{} simulation has reached convergence with respect to this observable. We revisit the tension between \Fable{} predictions and observations in Section~\ref{sec:revisiting-tensions}, demonstrating how the models presented in Section~\ref{sec:astro_uncert} can further improve agreement between the predicted GWB amplitude and observations. Finally, we discuss our findings and conclude in Section~\ref{sec:conclusion}.

\section{Methods} \label{sec:methods}

\subsection{The \Fable{} simulation} \label{sec:fable}

In this work, we use the \Fable{} simulation suite to study the predicted GWB from a population of merging SMBHs. \Fable{} is a large-volume cosmological hydrodynamical simulation run with the \textsc{Arepo} code, which tracks gas and collisionless dark matter (DM), star and BH particles. The simulation is evolved to $z=0$ from initial conditions consistent with a Planck cosmology ($\Omega_\Lambda = 0.6935$, $\Omega_M = 0.3065$, $\Omega_b = 0.0483$, $\sigma_8 = 0.8154$, $n_s = 0.9681$, $H_0 = 67.9$~km s$^{-1}$~Mpc$^{-1} = h\,\times \,100$~km s$^{-1}$~Mpc$^{-1}$). The largest simulation box has a volume of $(100 \, \mathrm{cMpc} \, h^{-1})^3$ with $1280^3$ DM particles of mass $m_{\rm DM} = 3.4\times10^7\,h^{-1}$~\Msun{} and an initial number of $1280^3$ gas resolution elements with a target mass of $\overline{m}_b = 6.4\times 10 ^6 \, h^{-1}$~\Msun{}, from which stars and BHs form. The gravitational softening length is set to $2.393 \, h^{-1}$~kpc in physical coordinates below $z = 5$ and in comoving coordinates for higher redshifts.

BHs are seeded in DM haloes above $5\times 10^{10} \, h^{-1}$~\Msun{} with an initial mass of $10^5 \, h^{-1}$~\Msun{} by converting the densest gas cell in the halo into a BH particle. To ensure BHs remain near the centres of galaxies, a repositioning scheme relocates each BH at every time step to the local gravitational potential minimum within its smoothing length.

BHs in \Fable{} grow through mergers and accretion. Accretion happens at an Eddington-limited Bondi-Hoyle-Lyttleton-like rate. Two BH particles are considered to merge when they come within a particle smoothing length of each other. This is an adaptive length that is used to estimate `local' hydrodynamic properties and is typically of the order of a few kiloparsecs (kpc). The distribution of these merger separations is shown in Fig.~8 of \citetalias{buttigieg_2025}. Because of the repositioning technique used, BH particle velocities are not tracked dynamically, and so two BH particles are merged independently of their relative velocity. The remnant BH inherits the sum of the progenitor masses, and merger events are recorded with higher time resolution than the simulation snapshots, including the scale factor and masses of both BHs at the moment of merger. This catalogue of `numerical BH mergers' forms the basis of our analysis of the BH merger population relevant for GWB predictions. We keep all merger events from this catalogue in which both BH particles have a mass $\geq10^6$~\Msun{} at the time of merger, to avoid results which sensitively depend on the seeding prescription. 

More details about the original \Fable{} simulations with a $(40 \, \mathrm{cMpc} \, h^{-1})^3$ volume are given by \citet{Henden_2018, Henden_2019, Henden_2019b} and its extension to the larger $(100 \, \mathrm{cMpc} \, h^{-1})^3$ cosmological volume by \citet{bigwood_2025}. We refer to these two cosmological boxes as \Fable{}-40 and \Fable{}-100, respectively, for the rest of this paper. We note here that the two boxes have the exact same mass and spatial resolution and differ only in their simulated volume. The BH population is investigated and validated against observations in our recent work~\citep[][hereafter SB25]{buttigieg_2025}. In this work, we use the merger catalogue from the larger \Fable{} simulation unless stated otherwise.

As discussed in~\citetalias{buttigieg_2025}, BH mergers in \Fable{}, and other simulations employing BH repositioning techniques, suffer from inaccurate BH dynamics that can lead to premature BH mergers relative to the merger timescales of their host galaxies. During galaxy interactions, BH particles may come within a smoothing length of one another and merge instantaneously in the simulation, even when the host galaxy merger is not yet complete. Additional artefacts arise when BH particles are repositioned during close encounters, causing them to lose association with their original host galaxies and resulting in incorrect merger histories. In~\citetalias{buttigieg_2025}, we address these issues by introducing `macrophysical' delays, which account for the time difference between the simulated BH merger and the host galaxy merger and precede any `microphysical' delays associated with unresolved hardening processes. These delays reduce both the BH merger rate and the predicted GWB amplitude (see Fig.~13 in \citetalias{buttigieg_2025}). In this work, our fiducial model uses the BH merger catalogue directly from \Fable{} without these corrections. We then reintroduce the macrophysical delays to construct a more physically motivated BH population and quantify the impact of these numerical artefacts on the predicted GWB amplitude and the consistency with current PTA measurements.

\subsection{Generating a GWB signal} \label{sec:methods_holodeck}

As discussed in Section~\ref{sec:fable}, the relevant output from a cosmological simulation like \Fable{} for predicting the GWB is a catalogue of discrete BH merger events. Each entry specifies the masses of the BHs, the redshift of the numerical merger event, and the separation at which they are merged in the simulation, which serves as the initial separation for subsequent binary evolution. To investigate astrophysical uncertainties, this catalogue of discrete events is post-processed to modify selected properties of these binaries to account for the possible BH pair/binary evolution on unresolved scales (see Section~\ref{sec:astro_uncert} for more details). To generate predictions of the GWB from such discrete populations, we use the \texttt{holodeck} code introduced by~\citet{Kelley_2017}\footnote{The \texttt{holodeck} code can be accessed via:
\url{https://github.com/nanograv/holodeck}.}

The method implemented in \texttt{holodeck} is described in detail by \citet{chen_2025} (see their Appendix~B). In summary, the characteristic strain in a frequency bin $[f, f+\Delta f]$ for the $i$-th realisation is given by
\begin{equation}
h_{c,i}^2(f,\Delta f) = \frac{f}{\Delta f}\sum_j \Lambda_{ij}h_{s,j}^2(f_r) \Bigg\rvert_{f_r = f(1+z)}\,,
\end{equation}
where the sum runs over all binaries $j$, each contributing an angle- and polarisation-averaged strain $h_s(f_r)$. The binaries are assumed to be on circular orbits (although see Sec.~\ref{sec:spin} for a discussion on eccentricity), and the source-frame frequency $f_r$ is related to the detector-frame frequency $f$ via $f_r = f(1+z)$. Each binary is weighted by a Poisson factor $\Lambda_{ij}$, which accounts for the comoving volume from which its GW signal can reach the Earth and for the binary’s effective lifetime. This stochastic weighting captures the discrete nature of the underlying population and the associated cosmic variance. By generating multiple realisations of the signal, we can quantify the intrinsic theoretical uncertainty in the predicted GWB amplitude, which is essential to robustly assess the level of tension between models and data, as will be discussed in Section~\ref{sec:tension}.

In this work, we evaluate the GWB on the frequency grid used in the NANOGrav free-spectrum analysis $f_{\mathrm{free}^{(i)}} = [0.06, 0.12, 0.19, 0.25, 0.31, ..., 1.87] \, \mathrm{yr}^{-1}$. Unless otherwise stated, any value quoted at a given frequency corresponds to the closest bin on this evaluation grid.

BH particles in \Fable{} are numerically merged while still being separated by kpc scales~\citep[see Section~\ref{sec:fable} and][]{Kelley_2016, buttigieg_2025}. Unresolved sub-kiloparsec hardening processes, which introduce additional microphysical delays and are necessary to generate a GWB signal, still have large associated uncertainties~\citep[see][for a review]{colpi_2014}. In this work, we adopt the simplified model of \citet{agazie_2023}, in which all binaries are assigned a fixed total hardening timescale $\tau_f$, which we set equal to 1~Gyr for our fiducial model unless explicitly stated. We explore the impact of varying $\tau_f$ in Section~\ref{sec:hardening_time}. 

The GWB generated using this method from the \Fable{} simulation, together with that resulting from the model introduced in \citetalias{buttigieg_2025}, can be seen in the top-left panel of Fig.~\ref{fig:gwb_comparison}.

\section{Is there a GWB tension?}
\label{sec:tension}

\begin{figure*}
    \centering
    \includegraphics[width = 2\columnwidth]{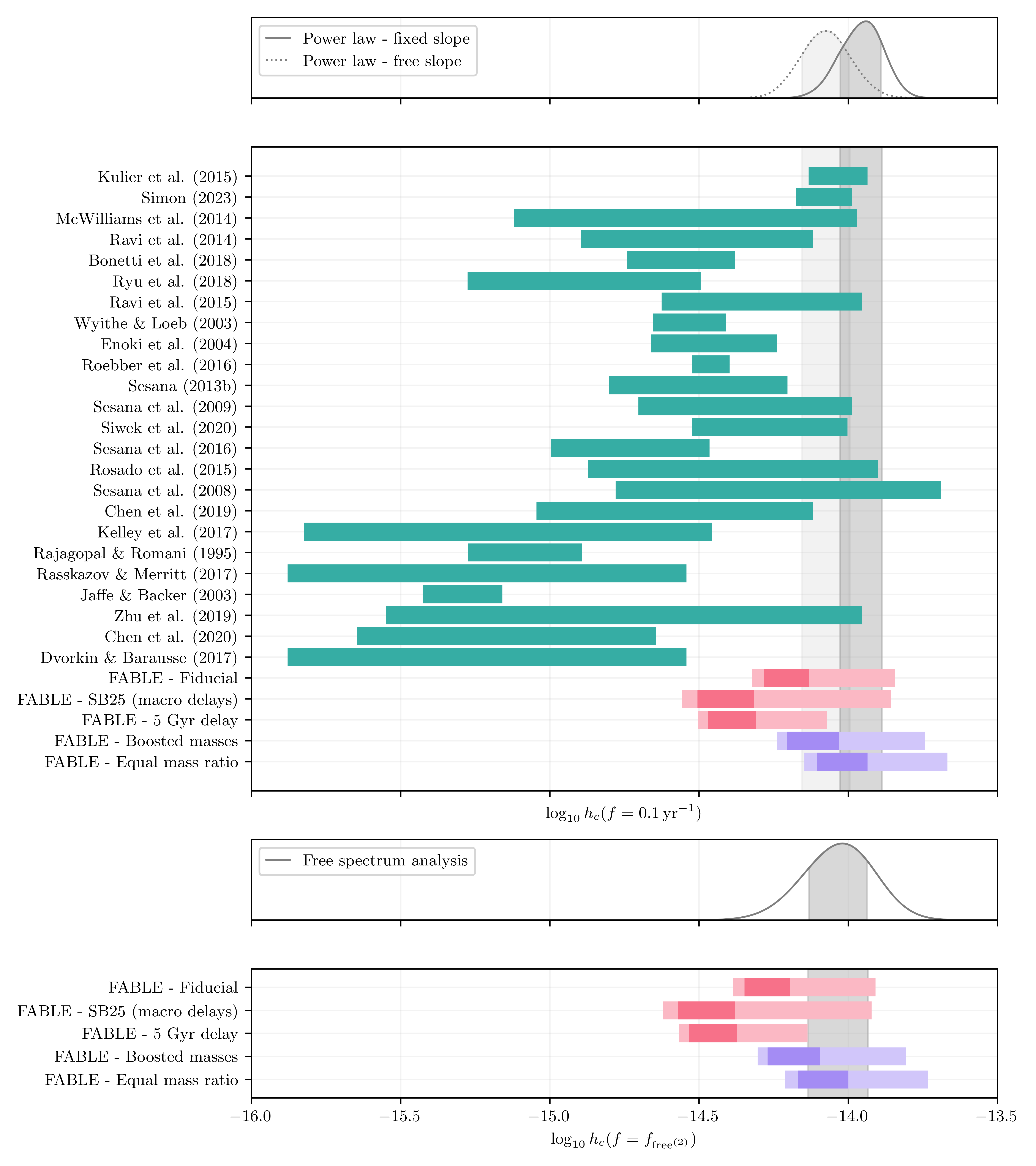}
    \caption{Adaptation of the right-hand-side plot in Figure A1 by \citet{agazie_2023}. The top panel shows the measured NANOGrav amplitude assuming a power-law with a fixed slope (solid) and a free slope (dotted) in their 15-year dataset. The grey shaded regions indicate the 68\% confidence interval of each distribution. These are then compared to theoretical amplitude predictions in the second panel. The models in teal show predictions from previous works, the ones in pink are direct predictions from the \Fable{} simulation, while the models in purple are more optimistic modifications to the population of SMBHBs in \Fable{} as discussed in Sections~\ref{sec:astro_uncert} and \ref{sec:volume}. We emphasize that literature predictions in teal were made prior to the release of PTA results; \Fable{} results shown in pink are postdictions that were obtained independently of the PTA measurements; and the more optimistic \Fable{} results shown in purple are postdictions constructed with the observed PTA signal in mind. All \Fable{} models assume a 1~Gyr hardening time with the exception of the \Fable{} 5~Gyr delay model. The teal and dark pink/purple horizontal bars indicate the 16th - 84th percentile uncertainty regions ($1\sigma$) for each prediction, while the light pink/purple bars show the 2.5 - 97.5 percentiles ($2\sigma$). The bottom two panels show the same comparison but for the second lowest frequency bin in the free-spectrum analysis of the same dataset, $f_{\mathrm{free}^{(2)}} = 0.12 \, \mathrm{yr}^{-1}$. In the top panel, in order to compare different models, all amplitude predictions are shifted to a common frequency of $f = 0.1\,\mathrm{yr}^{-1}$ assuming the expected $h_c \propto f^{-2/3}$ power-law scaling. The literature values are taken from \citet{rajagopal_1995,jaffe_2003,wyithe_2003,enoki_2004, sesana_2008, sesana_2009,sesana_2013b,mcwilliams_2014, ravi_2014, kulier_2015, ravi_2015, rosado_2015, roebber_2016,sesana_2016, dvorkin_2017,Kelley_2017,rasskazov_2017,bonetti_2018,ryu_2018,chen_2019,zhu_2019,chen_2020,siwek_2020, simon_2023}}. 
    \label{fig:model_comparison}
\end{figure*}

Fig.~\ref{fig:model_comparison} reproduces, with modifications, the comparison presented by \citet{agazie_2023} between the GWB amplitude inferred from the NANOGrav 15-year dataset and a range of theoretical predictions (teal bars), at a reference frequency of $f = 0.1 \, \mathrm{yr}^{-1}$. While some models are broadly consistent with the observed signal, many tend to predict amplitudes somewhat lower than those inferred from PTA observations. Subsequent studies have shown that the observed amplitude can be reproduced with relatively modest adjustments to assumptions about SMBH populations once the signal strength is known~\citep[see, for example,][]{bonoli_2025}. The pink bars show results from the \Fable{} cosmological simulation, described in Section~\ref{sec:fable}. Our \Fable{} models shown here assume a fiducial hardening timescale of 1~Gyr for all binaries to account for unresolved sub-grid physics governing SMBHB orbital evolution, as well as a more conservative 5~Gyr case. They also incorporate macrophysical delay prescriptions that correct for premature SMBH mergers caused by numerical limitations~\citepalias{buttigieg_2025}. The purple bars highlight more optimistic scenarios that can enhance the predicted amplitude, including modifications to the SMBHB population motivated by EM observations and astrophysical uncertainties, as discussed further in Sections~\ref{sec:astro_uncert} and~\ref{sec:volume}. The grey distributions shown in the top and second-to-bottom panels in Fig.~\ref{fig:model_comparison} demonstrate that the inferred strain amplitude measured by PTAs depends on the adopted data analysis method: assuming a fixed or free power-law slope, or using a free-spectrum analysis that allows each frequency bin to vary independently. Different assumptions thus yield varying amplitudes and, consequently, different apparent levels of tension with theoretical predictions.

While PTAs have now reported strong evidence for a common-spectrum process with the spatial correlations expected from a GWB, it is important to note that the data itself is not free from potential biases. For example, \citet{ferranti_2025} show that the recovered spectral properties of the background can be biased by the limited frequency coverage of present PTAs, while \citet{valtolina_2024} demonstrate how realistic noise and sampling effects can mislead standard detection analyses. More recently, \citet{crisostomi_2025, quelquejay_2025} point out that a finite observation window could lead to inter-frequency correlations that might be biasing spectral inferences towards larger amplitudes and shallower slopes. Other studies also caution that un-modelled noise sources and uninformative noise priors may bias the inferred signal properties~\citep{Zic_2022, Goncharov_2025, vanHaasteren_2025}. Together, these results emphasise the need for care when interpreting the statistical significance or detailed spectral shape of the signal at this early stage. In this work, however, our focus is on the astrophysical interpretation, and we therefore examine the agreement between theoretical predictions and the data constraints inferred under the standard NANOGrav analysis, as has been done in previous works by, for example, \citet{Sato-Polito_2024, Sato-Polito_2025}.

Before quantifying the tension between the \Fable{} cosmological simulation and the NANOGrav 15-year dataset, as shown in Fig.~\ref{fig:model_comparison}, in more detail, the first step is to define precisely what we are comparing to. As mentioned above, pulsar timing residuals can be analysed in different ways depending on the assumptions made about the form of the GW signal:
\begin{enumerate}
    \item \textbf{Power-law analysis:}
    One approach is to assume that the characteristic strain follows a power-law form, $h_c(f) = A_\mathrm{yr}(f/\mathrm{yr}^{-1})^{-\alpha}$, where $A_\mathrm{yr}$ is the GWB amplitude at $f = 1 \, \mathrm{yr}^{-1}$. The slope $\alpha$ may be fixed to the theoretically expected value of $2/3$, which is appropriate for a population of circular, GW-driven binaries~\citep{phinney_2001}. Otherwise, it can be treated as a free parameter to allow for potential deviations from this scaling. In both cases, the signal is assumed to follow the HD angular correlation pattern, which isolates the quadrupolar spatial correlations expected from a GWB~\citep{Agazie_2023b, agazie_2023}. Both variants of this analysis are shown in the top panel of Fig.~\ref{fig:model_comparison} at $f = 0.1 \, \mathrm{yr}^{-1}$.
    \item \textbf{Free-spectrum analysis:}  
    Alternatively, the amplitude of the signal in each frequency bin can be treated as a free parameter, yielding posteriors for the strain in each frequency bin. Different variants of the free-spectrum analysis exist depending on which correlated noise components are included. In this work, we adopt the HD-correlated free-spectrum model and compare different variants of the free-spectrum analysis together with their impact on the inferred tension values in Appendix~\ref{appendix:free_spectrum}.
\end{enumerate}

\begin{figure*}
    \centering
    \includegraphics[width = 2\columnwidth]{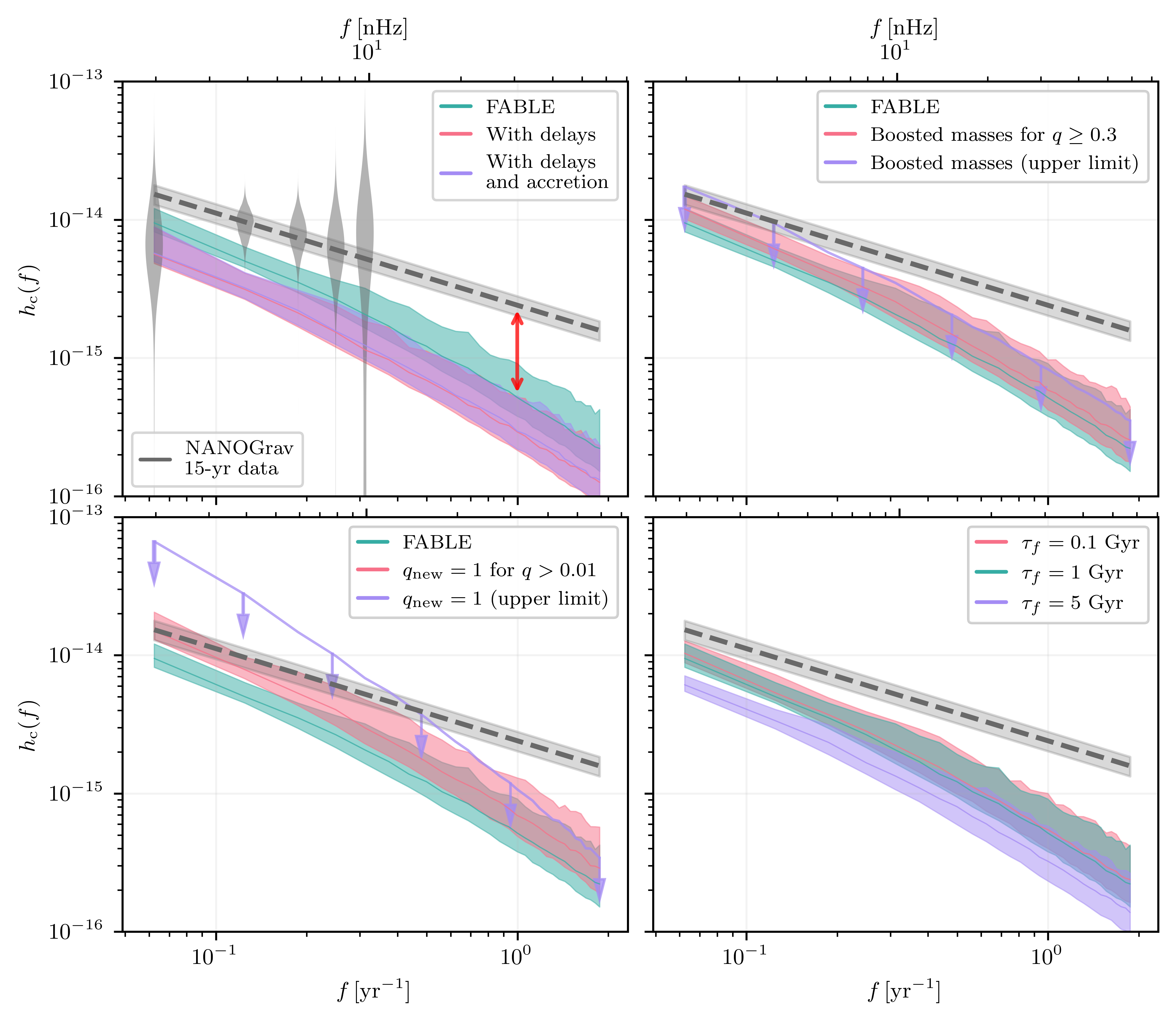}
    \caption{Predictions for the GWB from the \Fable{} simulations under various post-processing modifications. In all panels, the coloured lines represent theoretical predictions generated from 1000 realisations of the discrete binary population. Solid lines show the median, and the shaded regions indicate the 68\% confidence interval (16th–84th percentiles or $1\sigma$) for each model. The grey dashed line and shaded region correspond to the median and 68\% confidence interval from the NANOGrav power-law analysis with a fixed slope, i.e. $h_c \propto f^{-2/3}$. Grey violins (shown only in the top-left panel) display the Hellings-Downs correlated free-spectrum posteriors for the first five frequency bins. The plotted frequency range corresponds to the sensitivity window of pulsar timing arrays (PTAs) after 20 years of observations. The red arrow highlights the tension between the fiducial \Fable{} prediction and the power-law amplitude at a `reference' frequency of $1\,\mathrm{yr}^{-1}$.
    The panels show the impact of different model assumptions: top left, the GWB prediction from unmodified \Fable{} compared to the case including macrophysical delays (recreation of Fig.~13 in \citetalias{buttigieg_2025}); top right, the effect of boosting BH masses by a factor $f_b = 5$ for merger events at $z > z_t = 1$; bottom left, the effect of equalising mass ratios; bottom right, the effect of varying the fixed hardening time $\tau_f$ of binaries (0.1, 1, and 5~Gyr). In the bottom left panel, the upper limit is shown by setting all mass ratios to one (purple line with arrows), while a less extreme scenario adjusts only binaries with original mass ratio $q>0.01$ to $q_\mathrm{new} = 1$. Similarly, in the top right panel, the upper limit is set by boosting the masses in all merger events, while the less extreme scenario only boosts the masses of BHs involved in mergers with $q \geq 0.3$. Unless otherwise stated, all models assume a fixed 1~Gyr hardening time and use the unmodified \Fable{} population as the original population before the respective modification is applied.}
    \label{fig:gwb_comparison}
\end{figure*}

Several methods have been proposed in the literature to quantify discrepancies between posterior distributions, which in this context correspond to those derived from the data and from simulations~\citep[see][for a comparison of methods]{charnock_2017}. In this work, we choose the difference vector method, which we summarise in Appendix~\ref{appendix:difference_vector}. In order to compare our tension with the values found by \citet{Sato-Polito_2024}, we first apply this method at $f = 1 \, \mathrm{yr}^{-1}$, which is often treated as the `reference' frequency. Since the free-spectrum analysis is highly unconstraining in this bin, we use the power-law results with a fixed slope, while noting the limitations of this approach, as will be discussed below. When compared to the distribution of $h_c (f = 1\, \mathrm{yr}^{-1})$ generated from the \Fable{} population (obtained using the method discussed in Section~\ref{sec:methods_holodeck}), we find a discrepancy of $2.2\sigma$, which increases to $2.5\sigma$ when accounting for our macrophysical delays~\citepalias[see Section~\ref{sec:fable} and][]{buttigieg_2025}. These values are consistent with the $2$--$4.5\sigma$ discrepancies reported by \citet{Sato-Polito_2024}, although we do not reproduce the largest tensions found in their analysis (see their discussion of Fig.~4). This tension is highlighted by the red arrow in Fig.~\ref{fig:gwb_comparison}. Although we do not find a strong tension, the predicted amplitude remains noticeably lower than that inferred from PTA observations under the power-law assumption at this frequency, as is apparent from visual inspection of the GWB spectrum.

However, comparing theoretical models to observations using the power-law analysis at $f = 1\,\mathrm{yr}^{-1}$ has important limitations.
\begin{enumerate}
\item The power-law form $h_c \propto f^{-2/3}$ is derived under the assumption that the background is well described by a smooth superposition of many sources. In practice, there is a finite number of binaries contributing at these high frequencies since high-mass binaries merge before they can reach these frequencies, which can lead to deviations from a simple power-law scaling~\citep{sesana_2008}. Moreover, since the power-law model, by construction, imposes an additional constraint on the spectrum, it therefore yields tighter error bars than the free-spectrum analysis. However, these artificially narrow uncertainties should not be overinterpreted: if adopted directly, they would lead to systematically higher tension values, despite the fact that the underlying assumption of a smooth $h_c \propto f^{-2/3}$ scaling may not hold at high frequencies.

\item At the `reference' frequency of $1 \, \mathrm{yr}^{-1}$, the free-spectrum analysis provides no meaningful constraint, as this bin is mostly unconstrained by the data, giving only an upper limit on the strain. Any comparison at this frequency is therefore effectively an extrapolation from fits obtained at lower frequencies, where the signal-to-noise is higher. This means the apparent discrepancy at the `reference' frequency is not directly supported by the data and may not be representative of the true tension. This was already discussed by \citet{agazie_2023} and by the \citet{EPTA_2024}, and is why we plot the model comparison at the lower frequency of $0.1 \, \mathrm{yr}^{-1}$ in Fig.~\ref{fig:model_comparison}.

\end{enumerate}

With the current data, a better comparison between PTA observations and theoretical models would be to use the free-spectrum analysis posteriors in the lowest five frequency bins, where the data is most constraining~\citep[as also discussed by the][for example, in the discussion surrounding their Fig.~3]{EPTA_2024}. The free-spectrum analysis from the NANOGrav collaboration is shown as the grey violin plots in Fig.~\ref{fig:gwb_comparison}. In Appendix~\ref{appendix:difference_vector}, we describe in detail the procedure to obtain quantitative tension values from the free-spectrum posteriors and explore the impact of using different combinations of frequency bins on these tensions. 

The tension values we find using this approach for the \Fable{} simulation together with various astrophysical models described in Sections~\ref{sec:astro_uncert} and \ref{sec:volume} are discussed further in Section~\ref{sec:revisiting-tensions}. In summary, we find that the highest tension value for the unmodified \Fable{} simulation with a hardening timescale of 1~Gyr is $1.7\sigma$, which is found using the 2nd and 3rd lowest frequency bins (which are the most constraining), which then goes up to $2.5\sigma$ when incorporating our macrophysical delays or $2.4\sigma$ when prolonging the hardening timescale to 5~Gyr. Although the \Fable{} simulation is still under-predicting the amplitude of the GWB, when using the free-spectrum posteriors, the tension is not as significant as the power-law analysis seems to suggest. We note that these tension values depend sensitively on the assumed hardening timescale, increasing for longer values of $\tau_f$; a detailed exploration of this dependence is presented in Section~\ref{sec:hardening_time}.

\section{Astrophysical Uncertainties} \label{sec:astro_uncert}

Current observations provide limited constraints on the SMBH population and their merger rates, particularly at high redshift, while significant uncertainties remain in theoretical models of SMBH binary evolution, together leading to substantial uncertainty in the predicted GWB amplitude. Motivated by these uncertainties, we investigate how plausible variations in the underlying astrophysical assumptions affect the predicted GWB signal and whether they can increase its amplitude and improve agreement with current GWB observations. Specifically, we investigate the impact of: i) a different redshift evolution of the BH–host galaxy mass relation, ii) changing the merger mass ratios, iii) a wide range of possible binary hardening timescales and iv) BH spins and eccentricities.

\subsection{Boosted BH masses at high-$z$} \label{sec:mass_boosting}
Recent \textit{JWST} observations likely indicate the presence of overmassive BHs at high redshift relative to local scaling relations~\citep{Maiolino_2024, juodzbalis_2024, juodvzbalis_2025}. Observations at cosmic noon ($z \sim 1-3$) also identify SMBHs in low-mass host galaxies with BH masses two orders of magnitude larger than what the local scaling relations predict~\citep{mezcua_2024}. Motivated by these observations, we test the impact of boosting BH masses in mergers at early cosmic times. We implement a simple empirical model in which we increase the masses of BHs involved in merger events at $z > z_t$ by a factor $f_b$, applying the boost only to major mergers with mass ratio $q \geq 0.3$ to avoid excessively altering the BH mass function (see Appendix~\ref{appendix:mass_boosting}). BHs merging at $z < z_t$ or with $q < 0.3$ are left unchanged. Our goal is to explore how enhanced early SMBH growth might influence the amplitude of the GWB while keeping the local population largely consistent with present-day observations.

It is important to emphasise that this prescription is not intended to be a self-consistent physical model. Modifying BH masses at high redshift would alter the subsequent evolution of the systems, including their dynamics, hardening timescales and subsequent BH feedback. Moreover, any physical evolution in the BH-host galaxy mass relations is expected to occur relatively smoothly with redshift, on some characteristic timescale of the relevant physical process, rather than through a sharp transition at $z_t$ as implemented here. Nonetheless, this simplified model gives us insight into how such an evolving mass relation could impact the GWB. In particular, it allows us to gauge the magnitude of the BH mass `boost' required to affect PTA-band observables, identify which redshift ranges contribute most strongly, and build intuition for how early and efficient SMBH growth might reconcile simulations with current data. We also note that if BHs are indeed overmassive at high redshift, subsequent evolution may partially regulate this excess. At a given host galaxy mass, more massive BHs will lead to stronger feedback, thus self-regulating their accretion rates~\citep{koudmani_2021}, potentially leading to a very similar observed low-redshift BH population. In this sense, our model can be interpreted as redistributing BH growth to earlier cosmic times rather than altering the $z=0$ population.

\begin{figure}
    \centering
    \includegraphics[width=\columnwidth]{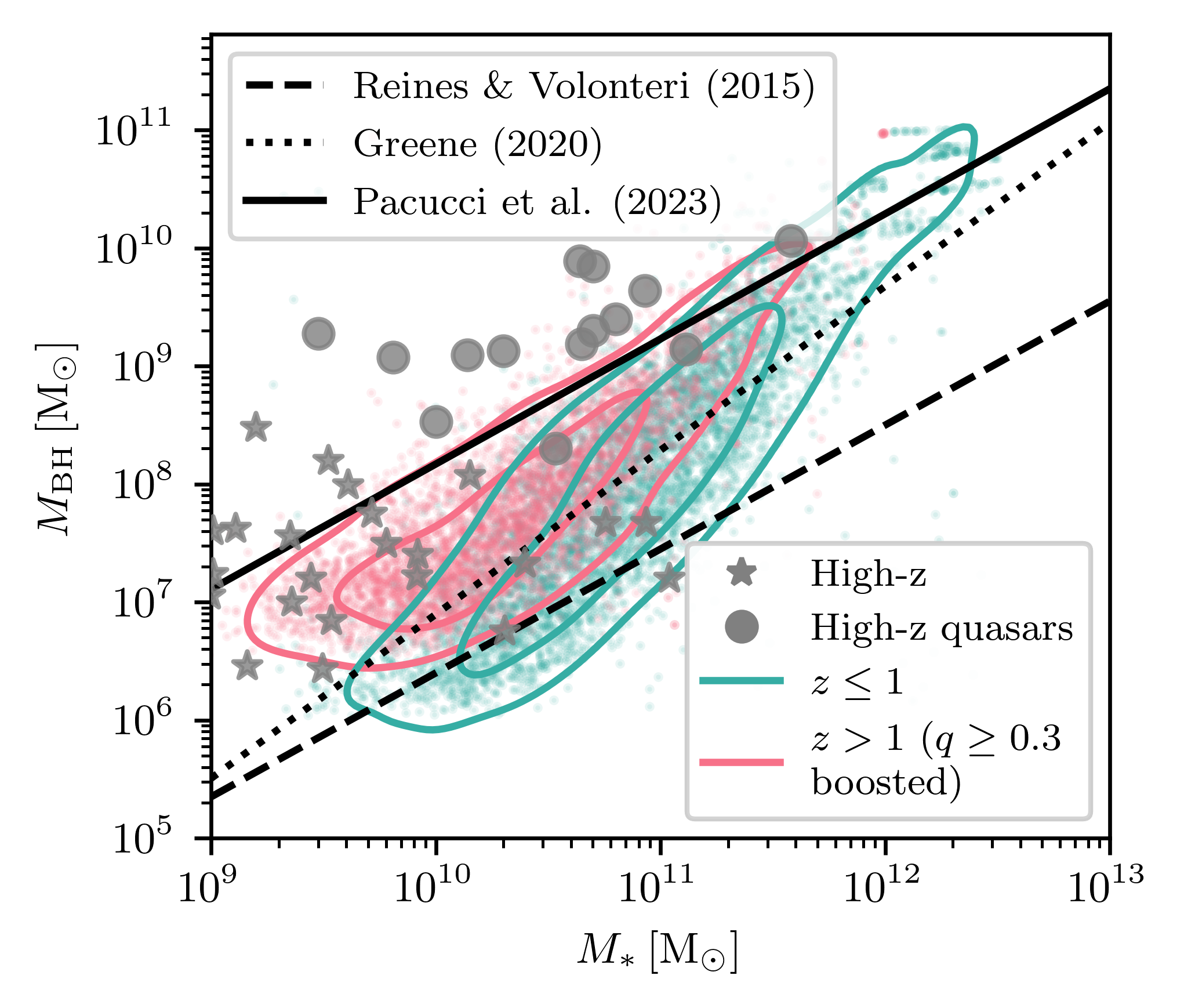}
    \caption{BH mass ($M_\mathrm{BH}$) vs. stellar mass within twice the half-mass radius of host subhalo ($M_{\mathrm{*}}$) for all primary BHs involved in SMBH mergers in \Fable{}. The scatter points show individual mergers while the contours indicate 50\% and 90\% of the population. The teal distribution shows all the merger events happening at $z \leq 1$ with their masses unchanged from the simulation. The pink distribution shows all merger events happening at $z > 1$ with the BH mass boosted by a factor of five if $q\geq 0.3$. The BH mass and host galaxy mass are taken at the snapshot immediately before the numerical BH merger in the simulation. The dashed and dotted black lines show the mean of the local relation by \citet{reines_2015} and \citet{greene_2020}, respectively. The solid black line shows the fit by \citet{Pacucci_2023} for \textit{JWST} objects at $z = 4 - 7$. The grey stars indicate high redshift observations from~\citet{harikane_2023,ubler_2023,maiolino_2024b, juodvzbalis_2025}.
    The grey dots are a compilation of high redshift quasar measurements from~\citet{Ding_2023, stone_2024, Yue_2024, marshall_2025}.}
    \label{fig:mass-relation}
\end{figure}

In the top-right panel of Fig.~\ref{fig:gwb_comparison}, we show results for a representative choice of parameters, $z_t = 1$ and $f_b = 5$. To illustrate the effect of our mass-ratio cut, we also show an upper bound corresponding to boosting all high-redshift BH masses regardless of merger mass ratio. We focus on this redshift threshold choice because boosting major mergers at higher redshifts ($z_t \gtrsim 2$) has a negligible impact on the predicted GWB amplitude, consistent with the fact that the signal is most sensitive to merger events at relatively low redshift ($z \lesssim 1.5$) as discussed by, for example, \citet{sesana_2013b, ravi_2015} and illustrated for the \Fable{} population later in Fig.~\ref{fig:contribution_grid}. 
The choice of $f_b$ is also intended as a representative choice, rather than being tuned directly to reproduce the observed GWB amplitude. Smaller values of $f_b$ produce small boosts in the GWB amplitude, while larger values provide a greater boost and broader agreement with observed high-redshift systems (see Fig.~\ref{fig:mass-relation}). However, such large values of $f_b$ require increasingly large amounts of synthetic BH growth that overshoot the growth of BHs in the \Fable{} simulation itself (see Appendix~\ref{appendix:mass_boosting} for more details). We therefore adopt $f_b = 5$ for the remainder of this work and refer the reader to simulations of different BH formation channels that directly investigate the BH mass--host galaxy mass relation at high redshifts~\citep[e.g.][]{bhowmick_2024, koudmani_2026}.

To investigate whether a boost for major mergers with $z_t = 1, \, f_b = 5$ would be consistent with current EM observations, we plot the $M_\mathrm{BH} - M_\star$ relation for all primary BHs involved in a merger event in Fig.~\ref{fig:mass-relation}, where $M_\star$ is the total stellar mass within twice the stellar half-mass radius hosting the BH with mass $M_\mathrm{BH}$. The teal distribution shows the BHs involved in a merger event occurring at $z \leq 1$, while the pink distribution shows the primary BH masses involved in a merger at $z > 1$, boosted by a factor of five if $q\geq 0.3$. These are compared to the mean local scaling relations by \citet{reines_2015} and \citet{greene_2020}, and the high-$z$ relation by \citet{Pacucci_2023}, which uses \textit{JWST} systems at $z = 4 - 7$. From this plot, it can be noted that the vast majority of `boosted' BH masses in the simulation lie between the local and high-$z$ relations, which is what we would expect if the $M_\mathrm{BH} - M_\star$ relation is evolving with redshift. 

Observations of high-redshift systems are shown by the grey stars in Fig.~\ref{fig:mass-relation}, taken from \citet{harikane_2023, ubler_2023, Maiolino_2024, juodvzbalis_2025}. A sample of high-$z$ quasars from \citet{Ding_2023, stone_2024, Yue_2024, marshall_2025} is also included, represented by grey circles. While some low-mass systems are consistent with both the local scaling relations and the unmodified \Fable{} distribution of primary BH masses, most high-redshift quasars agree more closely with the \citet{Pacucci_2023} relation and the boosted primary-mass distribution above $z_t = 1$. A subset of these quasars appear even more over-massive relative to this boosted relation. For visual clarity, error bars for these observed systems are omitted from Fig.~\ref{fig:mass-relation}; however, substantial uncertainties remain in both BH and host galaxy stellar masses. Several data points represent only upper limits on stellar mass, suggesting that the true BH-to-stellar-mass ratios may be even higher for parts of the quasar sample. BH mass estimates themselves are uncertain by up to $\sim$0.3~dex. Upcoming observations with \textit{JWST} and GRAVITY+ are expected to tighten constraints on both BH and host galaxy masses, potentially revealing more massive systems at $z \gtrsim 2$~\citep{gravity_2022, Abuter_2024}. 

Given the substantial scatter in current observational data, the possibility that the BH mergers contributing to the GWB at higher redshifts are more massive than those predicted by \Fable{} or similar cosmological simulations remains entirely plausible, and may help reconcile the slight underestimation of the GWB amplitude relative to PTA measurements. \citet{matt_2025} reach a similar conclusion when investigating the impact of a redshift-evolving $M_\mathrm{BH} - M_\mathrm{bulge}$ on the GWB.

\subsection{Mass ratios} \label{sec:mass_ratios}

The mass ratio, $q = M_2/M_1 \leq 1$, of each binary in the \Fable{} population used to generate the GWB signal is determined by considering individual BH masses when they are still at kpc-scale separations, owing to the limited spatial resolution of the simulation (see Section~\ref{sec:fable}). As these binaries evolve through various hardening mechanisms, their individual masses may change, potentially altering the mass ratio. For example, during gas-driven hardening, preferential accretion onto the secondary BH, an effect suggested by theoretical studies~\citep[e.g.][]{farris_2014, Gerosa_2015, siwek_2023}, could modify the mass ratio and consequently affect the resulting GWB.

To explore the possible impact of such evolution, we consider two empirical models. The first imposes an absolute upper limit on the amplitude boost attainable through variations in $q$, by redistributing the component masses such that every binary has $q=1$. The second, less extreme model, equalises the mass ratio only for binaries with an initial value of $q > 0.01$. This choice reflects the expectation that systems with highly unequal masses are unlikely to evolve toward equal mass ratios before binary coalescence. In such cases, the secondary BH would need to accrete an unrealistically large amount of gas to approach the mass of the primary, making significant mass ratio evolution highly unlikely.

This procedure should be regarded as a purely numerical experiment, in which mass is redistributed between the two BHs in each binary rather than added or removed. By construction, it preserves the total accreted mass of the system while altering only its distribution between the two components. This ensures that any change in the predicted GWB signal arises solely from modifications to the mass ratio, rather than from variations in the overall binary mass budget. In this way, our approach isolates the role of mass-ratio evolution in shaping the GWB, allowing us to assess its impact independently of other effects such as enhanced accretion or changes to the binary mass function.

The effect of these models is shown in the bottom-left panel of Fig.~\ref{fig:gwb_comparison}. Forcing all binaries to have $q = 1$ substantially increases the GWB amplitude, even exceeding the observed signal (purple line). The second approach, where the mass ratio is adjusted only for systems with $q > 0.01$, yields a GWB amplitude that is more consistent with the data without over-amplifying the signal (pink distribution). Importantly, this selective modification leaves the overall BH -- host galaxy mass relation (Fig.~\ref{fig:mass-relation}) largely unchanged, since the remnant BH masses remain the same as in the original simulation. However, even this latter model should be regarded as an upper limit, since we do not expect all binaries to evolve to $q \approx 1$ in practice, even under preferential accretion onto the secondary BH. Moreover, preferential accretion is expected to be significant in gas-rich mergers, whereas mergers occurring in gas-poor systems would largely retain their initial mass ratio as they do not accrete significantly during the hardening phase. The contrast between the two models indicates that the binaries with the most extreme mass ratios dominate the enhancement in the equal-mass case, as these systems contribute disproportionately to the boosted signal when their masses are equalised. While hydrodynamical simulations reveal a complex picture of mass-ratio evolution during gas hardening~\citep[see for example][]{bourne_2024}, our results suggest that any preferential accretion onto the secondary could help boost the GWB amplitude, a conclusion also found by~\citet{comerford_2025}.

The impact of the mass ratio re-distribution on the GWB strain can be seen through an analytical argument. For example, under certain simplifying assumptions, one finds that $h_c^2 (f) \propto \langle \eta \rangle$ where $\langle \eta \rangle$ is the average of the symmetric mass ratio $\eta = q/(1+q)^2$~\citep{phinney_2001, Sato-Polito_2024, sato_polito_2025b}. This implies that changes in the mass ratio distribution directly rescale the overall signal amplitude. In the most extreme case, when $q=1$ for every binary, $\langle \eta \rangle = 1/4$, which sets an upper limit to the `boost' in $h_c(f)$ purely from varying the distribution of mass ratios, corresponding to the upper limit shown in the bottom-left panel of Fig.~\ref{fig:gwb_comparison}.

\subsection{Hardening timescale}
\label{sec:hardening_time}

As discussed in Sections~\ref{sec:introduction} and~\ref{sec:tension}, one of the largest theoretical uncertainties in modelling SMBH binaries is the hardening timescale that shrinks their separation from the kpc scales at which they are merged in cosmological simulations down to coalescence. To assess the impact of this variable, we vary the total binary lifetime $\tau_f$ and compute the resulting GWB for $\tau_f = 0.1, \, 1,\, 5$~Gyr, as shown in the bottom-right panel of Fig.~\ref{fig:gwb_comparison}. Extending the lifetime to $5$~Gyr suppresses the amplitude by reducing the number of mergers occurring before $z=0$, while shortening it from $1$~Gyr to $0.1$~Gyr produces only a small amplification. Reducing $\tau_f$ to below $0.1$~Gyr, which is likely too short of a delay, would result in a suppression of the power at low frequencies as binaries spend less time at wide separations. These trends indicate that even under minimal delay assumptions, the GWB from \Fable{} mergers remains slightly below current measurements.

It is useful to place the hardening timescales adopted in our phenomenological model in the context of studies that explicitly model SMBHB evolution. These studies find timescales ranging from $\sim 0.1$~Gyr to several Gyr, with substantial variation depending on the properties of the host galaxy and binary. For example, the KETJU simulations find an average coalescence time of $\sim 200$~Myr after binary formation~\citep{rantala_2017}, while simulations that include nuclear star formation find merger times of $\sim 100$--$500$~Myr for disc mergers and $\sim 1$--$2$~Gyr for elliptical mergers~\citep{liao_2024}. RAMCOAL simulations find hardening timescales of $\lesssim 100$~Myr when viscous torques from a circumbinary disc are included~\citep{Li_2024}. Semi-analytic models coupled to cosmological simulations likewise predict a broad distribution of hardening timescales. For example, \citet{Kelley_2017} find binary lifetimes that can extend to several Gyr using the Illustris simulation. These results suggest that although efficient binary hardening with timescales $\lesssim 1$~Gyr is plausible, particularly for binaries in dense stellar or gaseous environments, substantial variation between systems motivates this exploration of a range of effective hardening timescales.

In our analysis of BH mass and mass ratio modifications so far, we adopted a relatively efficient hardening timescale of $1$~Gyr and neglected macrophysical delays to isolate the direct impact of these population changes~\citepalias{buttigieg_2025}. Here, we reintroduce the macrophysical delay correction to account for premature mergers of BHs due to the limited resolution, and investigate the joint effect of varying the hardening timescale together with the astrophysical population modifications described in Sections~\ref{sec:mass_boosting} and~\ref{sec:mass_ratios}.

\begin{figure}
    \centering
    \includegraphics[width=\columnwidth]{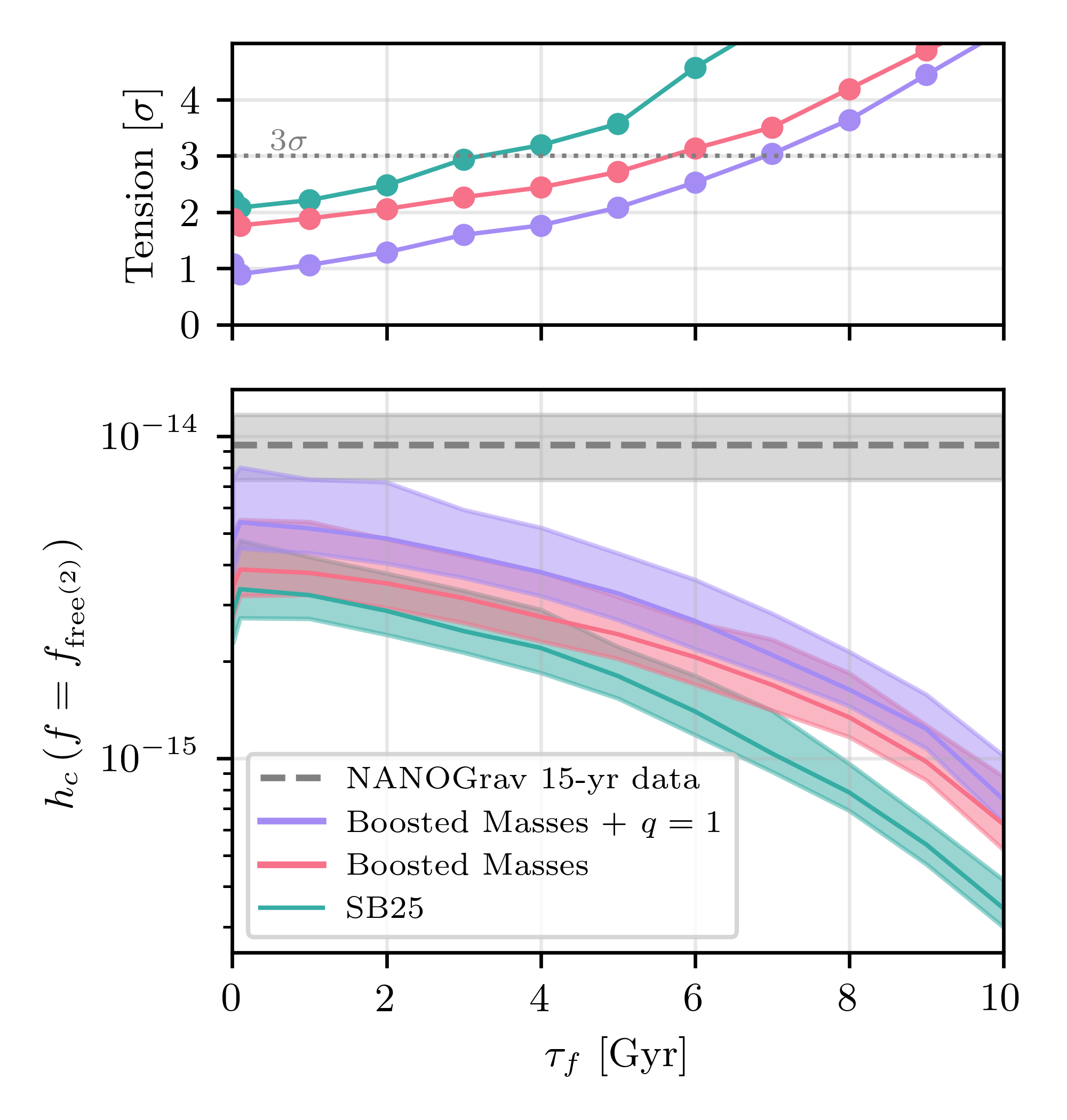}
    \caption{Bottom panel: predicted amplitude of the GWB at the second lowest frequency bin $f_{\mathrm{free}^{(2)}} = 0.12 \, \mathrm{yr}^{-1}$, shown as a function of the total binary hardening timescale $\tau_f$. The coloured lines represent theoretical predictions generated from 1000 realisations of the discrete binary population. Solid lines show the median, and the shaded regions indicate the 68\% confidence interval (16th–84th percentiles or $1\sigma$) for each model. The teal curve represents the \Fable{} population after applying the macrophysical delays described in \citetalias{buttigieg_2025}. The pink curve shows the same population (i.e.~with macrophysical delays) after boosting masses of major mergers above $z=1$ by a factor of five, while the purple curve additionally includes mass-ratio equalisation for systems with $q > 0.01$. The dashed grey line shows the median from the NANOGrav free-spectrum posterior with the shaded region showing the 1$\sigma$ confidence intervals. Top panel: the tension value of these models. These are calculated using only the second-lowest frequency bin from the free-spectrum analysis. To characterise the trend of tension with $\tau_f$, we employ a Gaussian approximation with an analytic p-value evaluation, rather than performing Monte Carlo sampling for each value of $\tau_f$. This approach captures the systematic behaviour of the tension while avoiding noise introduced by the Poisson sampling of \texttt{holodeck}, which can produce irregular behaviour in the tails that become relevant at high tension values (see left-hand panel of Fig.~\ref{fig:distribution}). We have verified that for $\tau_f \lesssim 5$~Gyr, the Gaussian approximation produces results consistent with full Monte Carlo sampling, justifying its use here to illustrate the overall trend without being dominated by sampling fluctuations.}
    \label{fig:timescales}
\end{figure}

Fig.~\ref{fig:timescales} shows the predicted GWB strain in the second-lowest frequency bin of the NANOGrav 15-year free-spectrum analysis as a function of $\tau_f$. The teal curve corresponds to the \Fable{} population after applying macrophysical delays~\citepalias{buttigieg_2025}. The pink curve includes the mass-boosting procedure for major mergers described in Section~\ref{sec:mass_boosting} for this delayed population of mergers, while the purple curve further incorporates mass-ratio equalisation for binaries with initial $q > 0.01$ (see Section~\ref{sec:mass_ratios}). The horizontal dashed line and shaded grey region show the NANOGrav 15-year measurement and its $1\sigma$ uncertainty band, respectively. The top panel then shows the tension values of these models with increasing hardening timescale, calculated using the method described in Appendix~\ref{appendix:difference_vector}, using only this second-lowest frequency bin. To guide the eye, and as a threshold above which we consider models to start showing significant tension with the data, we indicate a tension value of $3\sigma$ as the horizontal dotted line.

As the hardening timescale $\tau_f$ increases, the predicted GWB amplitude systematically decreases across all models, reflecting the reduced merger efficiency associated with prolonged binary evolution. At $\tau_f = 1$~Gyr, the baseline \citetalias{buttigieg_2025} model lies roughly $2\sigma$ below the NANOGrav measurement, whereas the most optimistic scenario shows a tension of around $1\sigma$. Incorporating astrophysical modifications increases the predicted amplitude, thereby improving consistency with the observed signal. The \citetalias{buttigieg_2025} model remains within $3\sigma$ agreement with the data for hardening times up to $\sim 3$~Gyr, while the boosted-mass model maintains consistency within $3\sigma$ for $\tau_f \lesssim 6$~Gyr. Equalising mass ratios extends this agreement to around $\tau_f \sim 7$~Gyr. Overall, even under conservative assumptions of inefficient hardening, the predicted amplitudes remain compatible with the observed GWB within current astrophysical uncertainties.

It is worth emphasising that these scenarios are not meant to represent a single physically realistic model of SMBH binary mergers in our Universe. Rather, they demonstrate the substantial astrophysical modelling freedom still permitted within observational constraints. Even with conservative merger rates and long inspiral times, the present GWB constraints are compatible with our models, which themselves remain consistent with EM observations of the local SMBH population.

\subsection{BH spins and eccentricity} \label{sec:spin}

BH spins affect the location of the innermost stable circular orbit (ISCO), which sets the separation, and thus the frequency, at merger. For non-spinning BHs, the ISCO lies at 3 Schwarzschild radii ($3 r_S$); maximally prograde spins (aligned with the orbit) shift it inward to $0.5r_S$, while maximally retrograde spins move it outward to $4.5r_S$~\citep{bardeen_1972}. This will impact the separation, and thus the frequency, until which the binary can emit GWs. However, merging BHs with different spin values produces no discernible change in the predicted GWB. The reason is that the GWB in the PTA frequency band is dominated by gravitational radiation from the long, low-frequency inspiral phase, when the two BHs are widely separated. The precise location of the ISCO, which affects only the final stages of coalescence, therefore has a negligible influence on the signal.

Note that BH spins may also influence the dynamics of SMBH binaries, and consequently their hardening timescales, for example by affecting the efficiency with which gaseous torques extract orbital angular momentum from the system~\citep{sayeb_2021, bourne_2024}. In our analysis, these effects are incorporated in the exploration of different hardening timescales described in Section~\ref{sec:hardening_time}. 

In addition, different spin orientations can produce a wide range of GW recoil kicks, with misaligned spins potentially able to eject remnant BHs from even the most massive host galaxies~\citep{merritt_2004, haiman_2004, volonteri_2006, blecha_2008,sijacki_2009, gerosa_2015b}. Recent work suggests that spin alignment may be inefficient even in gas-rich systems, with a non-negligible fraction of SMBH binaries retaining misaligned spins prior to merger and resulting in some BH remnants having a large recoil velocity~\citep[][but see \citealt{bourne_2024}]{sayeb_2021}. While \Fable{} does not self-consistently model recoil-driven ejections, such events could reduce the SMBH occupation fraction and merger rates. However, these spin-dependent effects are likely encompassed within the broader uncertainties associated with SMBH hardening and binary evolution~\citep[e.g.][]{bourne_2024,siwek_2024}, which already dominate the uncertainties in the predicted GWB amplitude considered in this work.

Eccentricity can also play an important role in the evolution of SMBHBs in the PTA frequency band. At large separations, eccentricity can shorten binary lifetimes, potentially increasing merger rates~\citep{Kelley_2017, Mannerkoski_2022}. Like the impact of BH spins on binary dynamics, this effect is captured phenomenologically by varying the hardening timescale in our models. However, if eccentricity persists into the PTA frequency band, it redistributes GW emission across higher harmonics and can therefore modify both the amplitude and spectral shape of the GWB~\citep{Sesana_2013, chen_2017, raidal_2024} potentially increasing the tension between predictions and observations.

Recent studies suggest that the residual eccentricity of SMBHBs in the PTA band may indeed be non-negligible. Gas-free simulations of galactic mergers with cosmological initial conditions find eccentricities of $e\sim0.2 - 0.8$ upon entry into the PTA band~\citep{gualandris_2022, fastidio_2024}, while simulations of consecutive major mergers find $e>0.85$ for all binaries in their sample~\citep{fastidio_2025}. When considering binaries embedded in circumbinary discs, \citet{siwek_2023b, siwek_2024} find that eccentricity evolves towards a mass-ratio-dependant equilibrium, reaching $\sim 0.5$ at higher mass ratios. These results suggest that, despite significant uncertainty in the expected residual eccentricity, non-negligible eccentricities may plausibly persist into the PTA frequency band. In this work, we focus on the effects of the SMBHB population and its associated astrophysical uncertainties. Moreover, the eccentricity of SMBHBs at PTA frequencies is difficult to constrain directly from the large-scale \Fable{} simulations, which do not resolve the binary evolution at the relevant sub-pc scales. We therefore adopt the simplifying assumption of circular binaries when computing the GWB.

\section{The high-mass end of the BHMF and finite-volume effects} \label{sec:volume}
Although the large $(100 \, h^{-1} \, \mathrm{cMpc})^3$ \Fable{} box represents a significant improvement over the original $(40 \, h^{-1} \, \mathrm{cMpc})^3$ box, this volume may still be insufficient to fully sample the most massive galaxies hosting SMBHs. Since the GWB is particularly sensitive to the high-mass end of the SMBH population, incomplete sampling of these rare systems could bias the predicted amplitude. In this section, we first compare the GWB predicted from the two \Fable{} volumes and then use synthetic SMBH populations drawn from the \Fable{} BHMF extended to higher BH masses, in volumes up to $(100 \, h^{-1} \, \mathrm{cMpc})^3$, to assess convergence of the predicted GWB signal in the PTA frequency range.

We test for this finite-volume effect by extending the \Fable{} BHMF using the halo mass function (HMF) from the large-volume, dark-matter-only Millennium-XXL (MXXL) simulation~\citep{angulo_2012}. MXXL is used here to inform the abundance of rare, massive haloes beyond those well sampled by \Fable{}, rather than to provide direct SMBH merger catalogues. We then construct multiple synthetic SMBH populations over a range of sampled volumes, up to the \Fable{}-100 volume, using a semi-analytical framework that incorporates the merger statistics measured in \Fable{}. This approach enables us to quantify realisation-to-realisation variance in the predicted GWB and to test whether rare, high-mass systems absent from one particular realisation of a given simulated volume could significantly enhance the background amplitude.

\subsection{Inferring the high-mass end of the black hole mass function} \label{sec:bhmf_extension}

\begin{figure}
    \centering
    \includegraphics[width = \columnwidth]{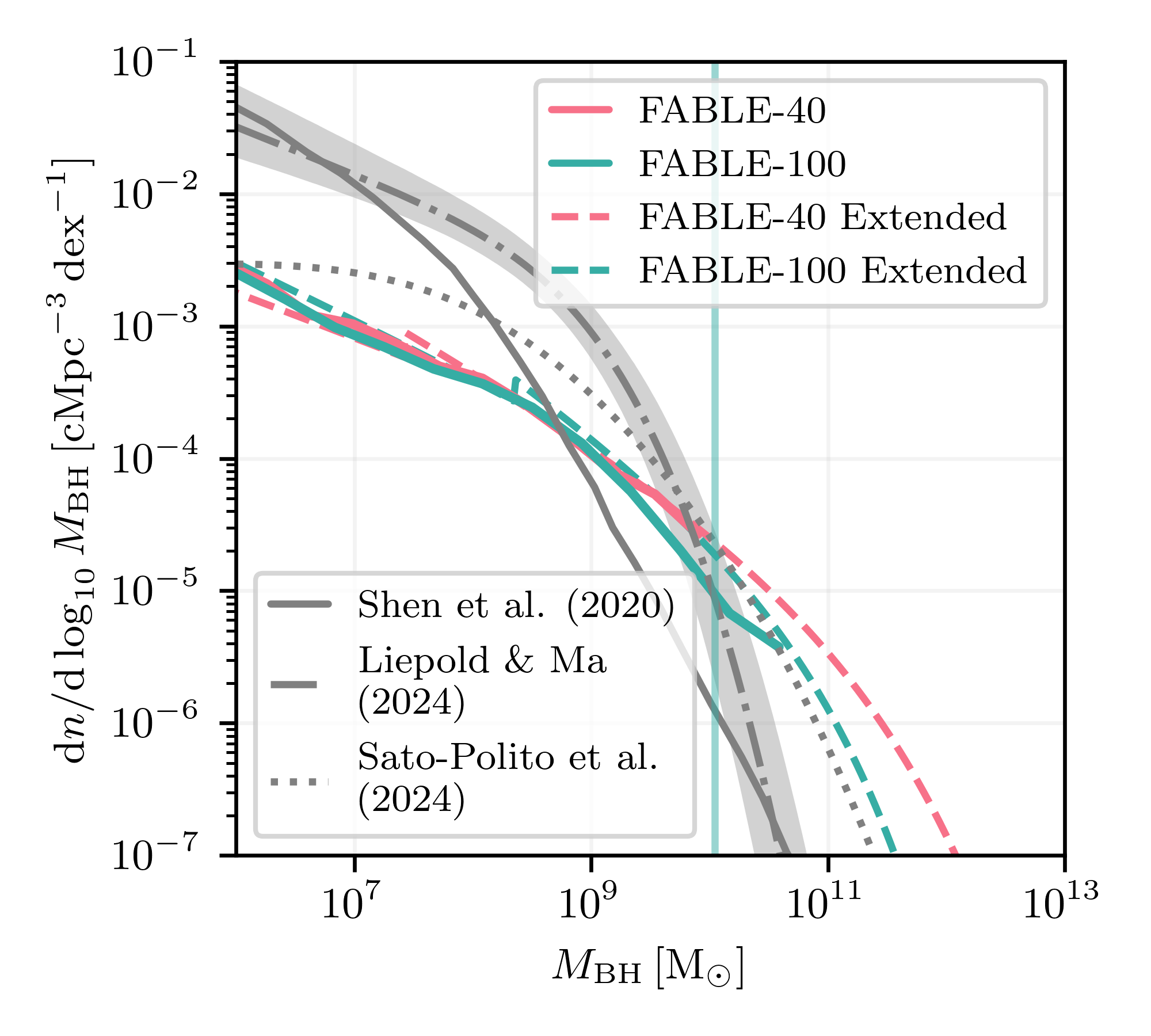}
    \caption{The BHMF for all BHs at $z=0$. The solid lines show the BHMF obtained directly from the simulation, with the pink line corresponding to the smaller \Fable{}-40 box, and the teal line to the larger \Fable{}-100. The dashed lines show the extended BHMF using the MXXL HMF as described in Appendix~\ref{sec:appendix-convergence}, with the corresponding colours showing which simulation box size was used for the extension. The grey solid line shows the BHMF calculated using the bolometric luminosity function of AGN in the local universe by \citet{Shen_2020} using their `deconvolution' method. The dashed-dotted grey line and shaded grey region show the median and 90\% confidence interval of the BHMF by \citet{liepold_2024}, obtained by combining their galaxy stellar mass function with the \citet{McConnell_2013} BH-stellar mass scaling relation. The dotted grey line shows the BHMF obtained by \citet{Sato-Polito_2024} to match the GWB strain. The vertical teal line indicates the maximum primary mass that contributes to the GWB at $f=0.1 \, \mathrm{yr}^{-1}$ for merger events with $q \ge 0.1$ in \Fable{}-100 (see Section~\ref{sec:synthetic_convergence}).}
    \label{fig:bhmf}
\end{figure}

The BHMFs measured directly from the \Fable{}-40 and \Fable{}-100 simulations at $z=0$ are shown in Fig.~\ref{fig:bhmf}. As expected, the larger \Fable{}-100 volume samples rarer and more massive BHs than the smaller \Fable{}-40 box, illustrating the strong dependence of the high-mass end of the BHMF on simulation volume. In the mass range where the two simulations overlap, their BHMFs are broadly consistent, indicating that the abundance of SMBHs is robustly captured at these masses. This agreement also persists at higher redshifts, as shown in Fig.~2 of \citetalias{buttigieg_2025}.

The extended BHMFs are constructed by fitting a broken power-law to the relation between SMBH mass and the mass of the host dark matter halo, and then applying this fit to the analytic MXXL HMF to obtain an extended \Fable{} BHMF~(details of the BHMF extension are provided in Appendix~\ref{sec:appendix-convergence}). In the mass range directly sampled by the simulations, the extended BHMF reproduces the simulated BHMF well, preserving both the normalisation and slope of the distribution. At higher masses, the method smoothly extrapolates the BHMF beyond the regime directly sampled by the simulations.

The BHMF extrapolated from the smaller \Fable{}-40 box predicts a higher abundance of SMBHs at the high-mass end compared to the extrapolation based on \Fable{}-100, a trend that is also observed at higher redshifts. This likely reflects the limited sampling of rare, massive systems in the smaller cosmological volume. In particular, the \Fable{}-40 simulation does not contain enough high-mass SMBHs to robustly constrain the slope of the BH mass–host halo mass relation at the high-mass end, which leads to an overestimation of the BH abundance when the relation is extrapolated. This should be interpreted as a limitation of applying this method to an insufficiently large cosmological volume, rather than evidence for a genuinely higher underlying BH abundance predicted by the simulation. By contrast, the larger \Fable{}-100 volume captures a greater number of massive systems and therefore provides a more reliable constraint on the high-mass behaviour of the BHMF. 

Nevertheless, the impact of uncertainties at the extreme high-mass end of the BHMF on PTA predictions, and the tension values calculated in this work, is limited. The vertical teal line in Fig.~\ref{fig:bhmf} marks the largest primary BH mass that contributes to the GWB at the most constraining PTA frequency, $f=0.1\,\mathrm{yr}^{-1}$, estimated from the \Fable{} merger population with $q\geq0.1$ assuming a representative hardening timescale of $1~\mathrm{Gyr}$. This provides a useful indication of the mass above which the detailed shape of the BHMF has little influence on the predicted GWB amplitude due to major massive mergers at frequencies where current PTA constraints are strongest. We return to this point in Section~\ref{sec:synthetic_convergence}.

In Fig.~\ref{fig:bhmf}, we compare these results to observationally inferred local BHMFs. The BHMF inferred by \citet{Shen_2020} from the bolometric luminosity function of AGN lies below both the directly measured and extended \Fable{} BHMFs at the high-mass end. The same is true when comparing to the BHMF of \citet{liepold_2024}, although the abundances of BHs match better towards the high-mass end when comparing to simulations. This difference is particularly important for the GWB predictions considered here, since the nanohertz GWB is dominated by mergers involving very massive SMBHs, especially at low frequencies. An enhanced abundance of massive SMBHs therefore leads to a larger predicted GWB amplitude, such that GWB predictions based on the \Fable{} BHMF are expected to lie above those based on EM-inferred BHMFs. This behaviour is not unique to \Fable{}. Fig.~13 of \citet{habouzit_2021} shows that several state-of-the-art cosmological simulations similarly predict an excess abundance of the most massive SMBHs relative to observationally inferred BHMFs at $z=0$. These simulations also find that the most massive BHs typically accrete at highly sub-Eddington rates and are therefore radiatively inefficient, suggesting that a significant fraction of the high-mass population may be faint or effectively dormant. Furthermore, theoretical arguments suggest that even if accreting efficiently, SMBHs at the extreme high-mass end of the BHMF may be difficult to observe because the ISCO exceeds the self-gravity radius of the disc, preventing the formation of a stable accretion disc~\citep{king_2016}. We return to this point in Sec.~\ref{sec:synthetic_convergence}. This interpretation is supported by the fact that, despite the apparent excess of massive BHs, the predicted bolometric luminosity function in \Fable{} remains in good agreement with observational constraints (see Fig.~3 in \citetalias{buttigieg_2025}), indicating that these systems contribute little to the observed AGN population.

It is also important to note that the high-mass end of the BHMF inferred from EM observations remains uncertain. As illustrated by Fig.~4 of \citet{liepold_2024}, different observationally inferred BHMFs exhibit substantial disagreement at $M_{\rm BH}\gtrsim10^{10}\,M_\odot$, reflecting uncertainties in the galaxy stellar mass function, velocity dispersion function, and the SMBH-host scaling relations. In addition, current galaxy surveys may not be sufficiently large to fully probe the abundance or properties of the rarest and most massive galaxies that host these extreme SMBHs~\citep{bernardi_2013, bernardi_2017, Sato-Polito_2024}. Because the high-mass tail of the galaxy stellar mass function is sensitive to both survey volume and systematic uncertainties in stellar-mass estimates, even modest incompleteness or mass underestimation can significantly affect the abundance of the most massive BHs obtained from these surveys.

We also compare our results in Fig.~\ref{fig:bhmf} to the BHMF obtained by \citet{Sato-Polito_2024}, who assume a broken power-law $M_\mathrm{BH}-\sigma$ relation convolved with the \citet{bernardi_2010} velocity dispersion function, adopting low-mass parameters from \citet{McConnell_2013} and calibrating the high-mass slope to match the GWB strain measured by PTAs. The high-mass end of this BHMF is in very good agreement with the BHMF extended from the \Fable{}-100 simulation. Since \citet{Sato-Polito_2024} infer a similarly enhanced abundance of massive SMBHs relative to EM-derived BHMFs, this agreement indicates that cosmological simulations independently reproduce the same excess population inferred from PTA observations. Interestingly, recent dynamical measurements have uncovered a growing population of ultramassive BHs with masses exceeding $10^{10}\,M_\odot$ that lie systematically above the canonical $M_\mathrm{BH}-\sigma$ relation~\citep{deNicola_2025}. Such systems are qualitatively more consistent with the steeper high-mass behaviour inferred by \citet{Sato-Polito_2024} and reproduced by the \Fable{}-100 extension. While the current sample of ultramassive BHs remains small, they provide evidence that the abundance of the most massive SMBHs may have been underestimated in previous EM-derived BHMFs.

Together, these results support the interpretation that the nanohertz GWB is sensitive to a population of extremely massive, weakly accreting SMBHs that is underrepresented in traditional EM-selected samples. This helps explain why \Fable{} GWB predictions show lower statistical tension with PTA measurements compared to BHMFs inferred from EM observations alone.

\subsection{Convergence tests using synthetic merger populations} \label{sec:synthetic_convergence}

The extrapolated BHMFs shown in Fig.~\ref{fig:bhmf} can be used to sample synthetic populations of BHs in a specified cosmological volume. A semi-analytical method based on the \Fable{} merger population (described in Appendix~\ref{sec:appendix-convergence}) can then be used to build synthetic merger catalogues. In this work we restrict these synthetic populations to the \Fable{}-100 volume and use them to quantify realisation-to-realisation variance arising from finite-volume effects. We do not use the extended BHMF to generate merger catalogues in substantially larger cosmological volumes. While the BHMF extension provides an estimate of the abundance of rare massive BHs beyond those directly sampled by \Fable{}, the merger statistics assigned to these systems are necessarily extrapolated from a very limited number of high-mass merger events in the simulation. Applying the method to much larger volumes could therefore artificially populate the high-mass end with large numbers of merger events whose merger rates are not directly constrained by \Fable{}. Restricting the analysis to the original simulation volume avoids this additional extrapolation while still allowing us to assess the impact of finite-volume sampling on the predicted GWB.

\begin{figure*}
    \centering
    \includegraphics[width = 2\columnwidth]{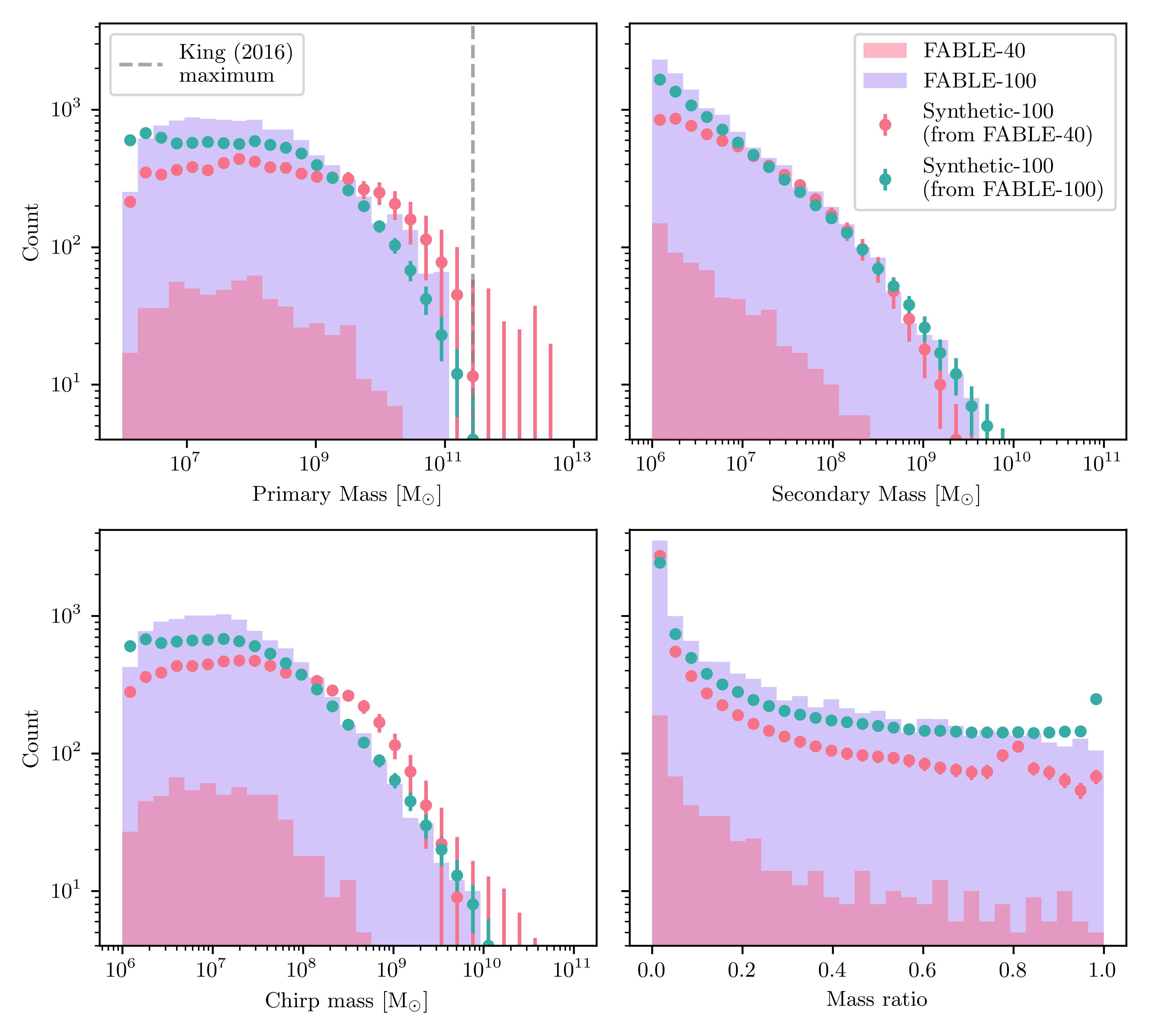}
    \caption{Distributions of merger event counts as a function of primary mass (top left), secondary mass (top right), chirp mass (bottom left), and mass ratio (bottom right). Filled histograms show the merger populations directly obtained from the smaller \Fable{}-40 volume (pink) and the larger \Fable{}-100 volume (purple). Points with error bars summarise the ensemble of 1000 synthetic merger catalogues generated using the method described in Appendix~\ref{sec:appendix-convergence} for a synthetic volume of $(100 \, h^{-1} \, \mathrm{cMpc})^3$. Teal points denote the median and standard deviation in each bin for synthetic populations informed by the \Fable{}-100 simulation, while pink points show the corresponding statistics for populations informed by \Fable{}-40 (but extended to the larger volume). The vertical dashed grey line in the top-left panel indicates the maximum BH mass attainable through luminous accretion as derived by \citet{king_2016}.}
    \label{fig:pop_histograms}
\end{figure*}

Fig.~\ref{fig:pop_histograms} compares the distributions of merger event counts as a function of primary mass, secondary mass, chirp mass, and mass ratio measured directly in the \Fable{} simulations with those obtained from these synthetic merger catalogues. The filled histograms show the merger populations in the \Fable{}-40 and \Fable{}-100 volumes. The points with error bars indicate the median and scatter across 1000 synthetic realisations for a volume of $(100\, h^{-1}\, \mathrm{cMpc})^3$, constructed using the method described in Appendix~\ref{sec:appendix-convergence}. This procedure is performed twice, once using the BHMF and merger statistics derived from each simulation, \Fable{}-40 and \Fable{}-100. Overall, we find good agreement between the simulations and the synthetic populations across the range where the simulations are well-sampled, indicating that our semi-analytical framework successfully reproduces the statistical properties of the merger population.

We note a slight overestimation of binaries with nearly equal mass ratios in the synthetic catalogues. This is driven by the fact that our \Fable{} merger catalogue imposes a cut requiring both the primary and secondary masses to be larger than $10^6$~\Msun{} (see Section~\ref{sec:fable}), which artificially biases sampled mass ratios toward equality when drawing BHs of around this mass from the BHMF. Since these low-mass systems contribute negligibly to the GWB and to avoid adding complications to the semi-analytical model, we do not correct for these artefacts.  

At the high-mass end, where direct simulation statistics are limited, the synthetic catalogues extend the distributions and provide an estimate for the expected contribution from rare massive systems sampled from the extended BHMF. We note that the synthetic populations generated using information from the \Fable{}-40 simulation include merger events with higher primary masses than those obtained using \Fable{}-100. This is due to the BHMF extended from \Fable{}-40 predicting a larger abundance of high-mass BHs, increasing the probability of sampling such systems, together with limited merger information at these high masses, which can lead to an overestimation of merger rates in this regime.

The maximum BH mass in \Fable{} is limited by the different simulation volumes, as can be seen in Fig.~\ref{fig:bhmf}, and the sub-grid modelling of galaxy–BH co-evolution. The extension procedure presented here effectively compensates for these resolution constraints while preserving the underlying physical scaling relations captured by \Fable{}. However, theoretical considerations impose an upper bound on luminous BH growth: beyond a certain mass, a BH can no longer sustain a stable accretion disc. As discussed by \citet{natarajan_2009} and \citet{king_2016}, this limit can reach $\sim 2.7\times10^{11}$~\Msun{} for maximally spinning, prograde gas accretion onto BHs. This limit is roughly three times larger than the most massive \Fable{} primary BH and is indicated by the vertical dashed line in Fig.~\ref{fig:pop_histograms}. BHs may still grow beyond this limit through mergers and possibly radiatively inefficient accretion, though such systems would no longer be observable as luminous AGN. We do not explicitly account for this theoretical upper limit on luminous growth in our extended populations, as the primary goal of this method is to determine whether a $(100\, h^{-1} \, \mathrm{cMpc})^3$ volume is sufficient to achieve convergence in the GWB amplitude.

\begin{figure*}
    \centering
    \includegraphics[width = 2\columnwidth]{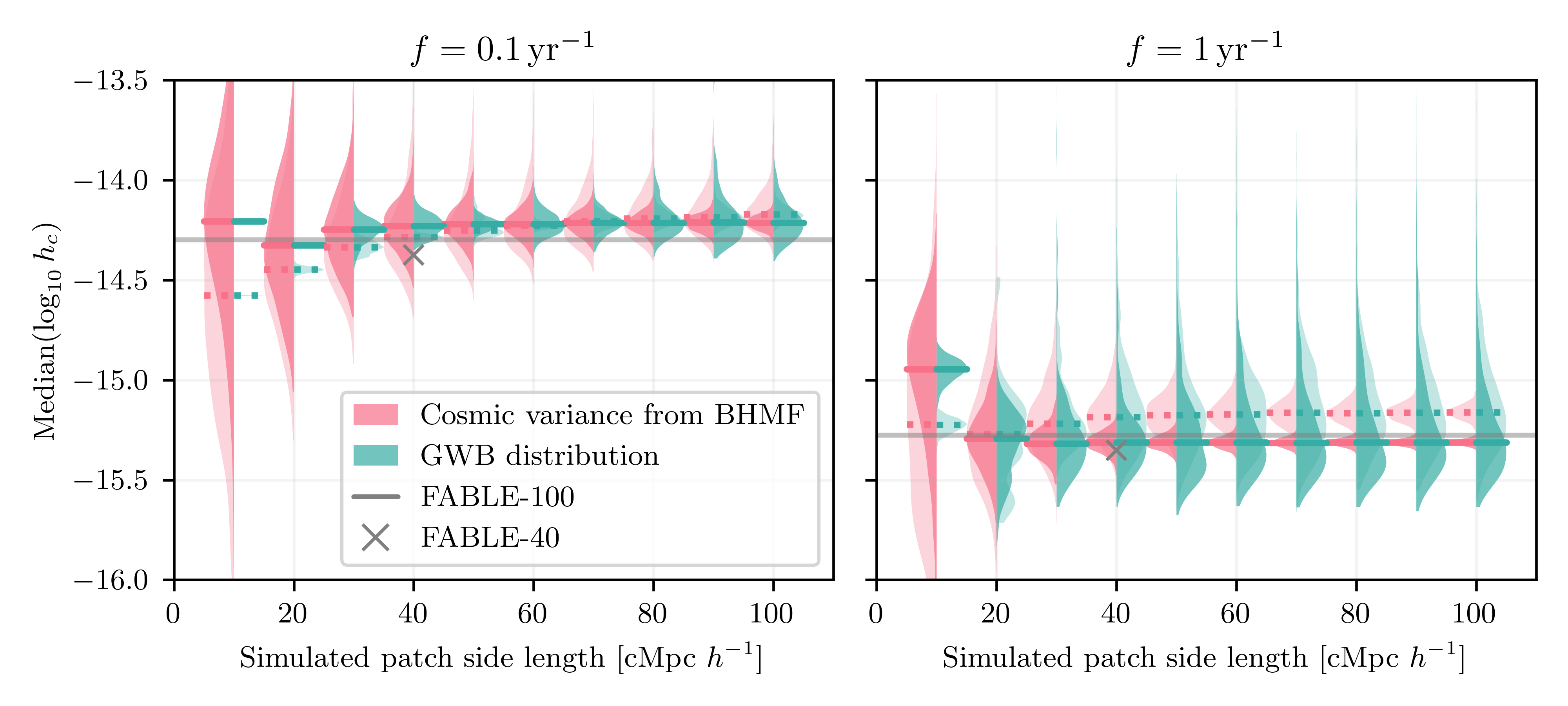}
    \caption{Distribution of the medians of $h_c$ from 1000 realisations of the GWB from 1000 synthetic populations as a function of the simulated patch size in cMpc $h^{-1}$ (pink violins on the left). The teal distributions on the right then show the distribution of the 1000 realisations for the population whose median lies closest to the median of all the populations, i.e., a representative population. The left panel shows these distributions at a frequency of $f = 0.1 \, \mathrm{yr}^{-1}$, while the right panel shows the higher frequency of $f = 1 \, \mathrm{yr}^{-1}$. The darker distributions correspond to the synthetic populations generated from the larger \Fable{}-100 box, while the lighter distributions show the results from \Fable{}-40. The horizontal solid/dotted lines show the medians of the distributions for the larger/smaller \Fable{} box. The median of 1000 realisations of the GWB generated from the actual simulated populations is shown in the solid grey horizontal line for \Fable{}-100, and the grey cross for the smaller \Fable{}-40.}
    \label{fig:convergence}
\end{figure*}

To assess convergence in the predicted GWB amplitude from \Fable{}, we generate synthetic merger populations for a range of cosmological volumes with side lengths between $10 \, h^{-1} \, \mathrm{cMpc}$ and $100 \, h^{-1} \, \mathrm{cMpc}$. For each volume, we construct 1000 independent populations by sampling from the underlying extended BHMF described in Section~\ref{sec:bhmf_extension}, thereby capturing cosmic variance arising from different realisations of the BH population. For each synthetic population, we use \texttt{holodeck} to generate 1000 realisations of the GWB following the procedure described in Section~\ref{sec:methods_holodeck}, and compute the median characteristic strain across these realisations. Fig.~\ref{fig:convergence} shows the distribution of these population medians when the synthetic catalogues are informed by the \Fable{}-40 and \Fable{}-100 simulations. These distributions (pink violins) are compared to the distribution of GWB realisations obtained for a single representative population (teal violins), which captures the stochastic variance associated with generating multiple GWBs from the same discrete merger catalogue using \texttt{holodeck}.

For small simulated volumes, cosmic variance from sampling the BHMF is large, leading to a wide spread in the median strain across different population realisations. In this regime, individual volumes may or may not contain rare, massive binaries that dominate the GWB, producing significant variation in the predicted amplitude. At the same time, the number of mergers within a given population is small, so the additional variance introduced by Poisson sampling within \texttt{holodeck} remains limited. As the simulated volume increases, the variance associated with BHMF sampling decreases because each realisation contains a more representative sampling of the underlying mass distribution, reducing the sensitivity to the presence or absence of rare high-mass systems. Conversely, the variance captured by \texttt{holodeck} grows as the number of contributing binaries increases, before approaching a plateau at large volumes. 

Within this framework, we consider convergence to be achieved when the variance arising from different BHMF realisations becomes smaller than the variance obtained by generating multiple GWB realisations from a single merger population. In other words, once the change in the median strain between different population realisations lies within the intrinsic stochastic uncertainty of the GWB calculation, increasing the simulated volume no longer significantly alters the prediction. The shape and width of the distributions depend on frequency, reflecting the fact that different frequency bands are sensitive to different parts of the SMBH mass spectrum, as will be discussed in more detail below. This is evident from the differences between the left and right panels of Fig.~\ref{fig:convergence}, indicating that convergence is frequency dependent.

For populations informed by \Fable{}-100, which provides more robust constraints at the high-mass end, convergence is reached at a box length of approximately $80\, h^{-1} \, \mathrm{cMpc}$ at the lower frequency of $f = (10 \, \mathrm{yr})^{-1}$, and at around $30\, h^{-1} \, \mathrm{cMpc}$ at $f = 1\, \mathrm{yr}^{-1}$. In contrast, when using the BHMF informed by \Fable{}-40, convergence is not fully achieved within volumes up to $100\, h^{-1} \, \mathrm{cMpc}$ at $f = (10 \, \mathrm{yr})^{-1}$, while it is reached at roughly $40\, h^{-1} \, \mathrm{cMpc}$ at $f = 1\, \mathrm{yr}^{-1}$. These differences are driven primarily by the high-mass behaviour of the extended BHMF (see Fig.~\ref{fig:bhmf}) and by differences in the underlying merger populations, although the qualitative trends are consistent across both simulations.

To isolate the impact of modelling assumptions in the semi-analytical framework (Appendix~\ref{sec:appendix-convergence}), we also generate a synthetic merger population by starting directly from the BH population in \Fable{}-100 rather than sampling from the extended BHMF. We find that the median strain obtained from 1000 GWB realisations of this synthetic population differs from that measured directly in \Fable{}-100 by $\Delta \log_{10} h_c (f = 0.1 \, \mathrm{yr}^{-1}) = 0.13$. This indicates that the offset between the strain measured in the simulation and the value toward which the synthetic populations converge is primarily due to small differences introduced by the semi-analytical modelling, rather than the inclusion of additional rare high-mass binaries. For this reason, we do not compute tension with PTA measurements and conclude that the absence of extremely massive systems in \Fable{}-100 is unlikely to significantly affect the predicted GWB amplitude.

\begin{figure*}
    \centering
    \includegraphics[width=2\columnwidth]{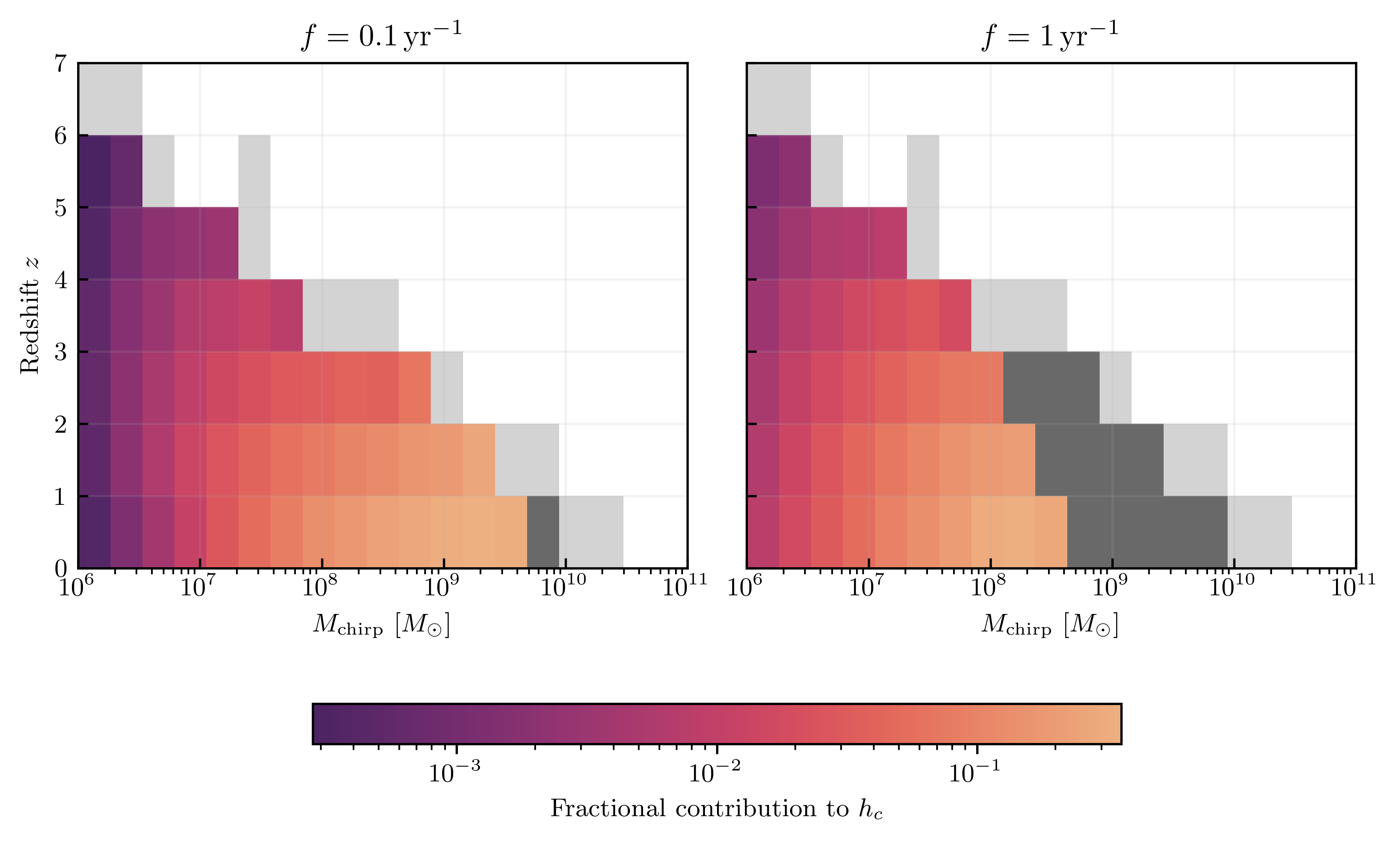}
    \caption{Fractional contribution to the characteristic strain $h_c$ as a function of chirp mass $M_\mathrm{chirp}$ and redshift $z$ for the \Fable{}-100 population. This is shown at a frequency of $f = 0.1 \, \mathrm{yr}^{-1}$ on the left and $f = 1 \, \mathrm{yr}^{-1}$ on the right. Light grey squares indicate bins with one to five merger events for which the GWB cannot be reliably computed using \texttt{holodeck}. Dark grey squares indicate a resulting GWB strain equal to zero from at least six merger events, while white squares indicate no merger events. A hardening timescale of $1~\mathrm{Gyr}$ is assumed throughout.}
    \label{fig:contribution_grid}
\end{figure*}

To assess which BH masses dominate the predicted GWB, and to understand why convergence in the GWB amplitude is frequency-dependant as highlighted in Fig.~\ref{fig:convergence}, we compute the fractional contribution to the characteristic strain, $h_c$, as a function of chirp mass and redshift for the \Fable{}-100 merger catalogue. Starting from the discrete SMBH merger catalogue, we bin mergers in $M_\mathrm{chirp} - z$ space and compute the median GWB from these sub-populations at two frequencies of $f = 0.1 \, \mathrm{yr}^{-1}$ and $f = 1 \, \mathrm{yr}^{-1}$. The resulting strain amplitudes are normalised by the value obtained from the full merger catalogue, yielding a fractional contribution to $h_c$. This is shown in Fig.~\ref{fig:contribution_grid}.

At both frequencies, mergers occurring at $z \lesssim 2$ dominate the signal. However, the masses that dominate the GWB depend on frequency. At $f = 0.1\,\mathrm{yr}^{-1}$, the strain is primarily sourced by binaries with $M_\mathrm{chirp} \sim 10^9\,\mathrm{M}_\odot$, while at $f = 1\,\mathrm{yr}^{-1}$ the dominant contribution shifts to $M_\mathrm{chirp} \sim 10^8\,\mathrm{M}_\odot$. This shift arises because binaries of different masses occupy different stages of their inspiral at a fixed observed frequency. Heavier binaries reach a given frequency later in their evolution, when the inspiral proceeds more rapidly. Consequently, although they generate stronger strain, they sweep through that frequency interval quickly and contribute relatively little power there. The GWB strain, therefore, peaks near the upper edge of the chirp-mass distribution before dropping sharply, where the increasing strain with mass is offset by the decreasing time spent emitting at that frequency.

This sensitivity of the GWB spectrum to different masses at different frequencies explains the frequency dependence of the volume convergence shown in Fig.~\ref{fig:convergence}. Smaller cosmological volumes adequately sample the masses dominating $h_c(f = 1\,\mathrm{yr}^{-1})$, whereas larger volumes are required to capture the rarer, high-mass binaries that dominate $h_c(f = 0.1\,\mathrm{yr}^{-1})$.

Fig.~\ref{fig:contribution_grid} visually shows which part of the high-mass tail of the BHMF matters for robust GWB predictions at the most constraining frequency. The highest-chirp-mass mergers at low redshift do not contribute appreciably to the background even at $f=0.1\,\mathrm{yr}^{-1}$. To connect this directly to the BHMF, we identify the most massive primary BH in merger events with $q\geq0.1$ that contributes at this frequency, finding $M_{\rm BH}=1.1\times10^{10}\,\Msun{}$, indicated by the vertical teal line in Fig.~\ref{fig:bhmf}. Thus, while larger volumes are needed to sample the binaries that dominate the low-frequency GWB, incompleteness in the extremely rare population above this mass is not expected to significantly affect the predicted amplitude at the PTA frequencies where observational constraints are strongest. At lower frequencies, where more massive binaries may become more relevant, current observational constraints on the background amplitude are substantially broader (see Fig.~\ref{fig:gwb_comparison}).

\section{Revisiting Tensions} \label{sec:revisiting-tensions}

In Section~\ref{sec:tension}, we discussed the caveats of directly comparing PTA data with theoretical predictions and introduced a statistical framework for quantifying the level of agreement between our models and the current best observational constraints. Here, we revisit the tension between the \Fable{} model and the NANOGrav 15-year dataset, extending this comparison to include the more optimistic models introduced in Sections~\ref{sec:astro_uncert}. We note that, as demonstrated in Section~\ref{sec:volume}, the predicted GWB strain is fully converged with respect to simulation volume for \Fable{}-100, ensuring that limited-volume effects do not impact the inferred tension values presented below.

Table~\ref{tab:tension} summarises the resulting tension values for five theoretical models: the fiducial \Fable{} model with a 1~Gyr hardening delay; a more conservative version with a 5~Gyr delay; \Fable{} including macrophysical delays~\citepalias{buttigieg_2025}; a boosted-mass model for major mergers with $z_t = 1$ and $f_b = 5$; and a model with equal mass ratios for binaries with original $q > 0.01$. All but the 5~Gyr delay case assume a 1~Gyr hardening timescale. For each model, we report the tension derived from a fixed-slope power-law spectrum ($\alpha = 2/3$) at the `reference' frequency of $1 \, \mathrm{yr}^{-1}$ which, as cautioned in Section~\ref{sec:tension}, can overstate discrepancies. We also show results from various combinations of free-spectrum frequency bins, which provide a more robust assessment of model–data consistency (see Appendix~\ref{appendix:difference_vector} for a detailed description of the method). Fig.~\ref{fig:model_comparison} provides a visual representation of the tension values in this table. 

The power-law tensions serve roughly as an upper bound on the disagreement between data and theory, since the method enforces a specific spectral shape that may not reflect the true form of the signal. Consequently, relying solely on the power-law analysis could lead to misleading conclusions about the level of tension. In contrast, the free-spectrum results reveal a more nuanced picture: the largest tensions occur when only the second and third frequency bins, i.e.,~those with the strongest constraining power, are included (see Fig.~\ref{fig:gwb_comparison}). Adding higher-frequency bins generally reduces the tension, indicating improved agreement between the model predictions and the data at those frequencies.

Although Table~\ref{tab:tension} presents results for the optimistic 1~Gyr hardening timescale, Fig.~\ref{fig:timescales} illustrates how increasing the hardening time systematically raises the tension values. Nevertheless, even for longer timescales, the models remain broadly consistent with the NANOGrav 15-year results as discussed in Section~\ref{sec:hardening_time}.

Overall, these results demonstrate that physically motivated SMBHB population models can reproduce the amplitude and spectral shape of the observed GWB within current uncertainties. While certain astrophysical scenarios (e.g., extended hardening times) can increase the apparent tension, none of the models considered here are decisively ruled out by present PTA data, underscoring the consistency between theoretical expectations and the nanohertz GWB. We note, however, that the relatively high abundance of massive BHs predicted by \Fable{} compared to that inferred from EM observations, in agreement with many other cosmological galaxy formation simulations~\citep{habouzit_2021}, likely contributes to the generally low tension values found in this work.

\begin{table*}
    \centering
    \small
    \caption{Tension values ($\sigma$) between different theoretical models and the NANOGrav 15-yr data, computed using both the power-law and free-spectrum analyses. These are calculated using the method described in Appendix~\ref{appendix:difference_vector} with a GWB generated from 100,000 realisations, using $1\times 10^6$ Monte Carlo points and running the analysis 50 times. Tension values are quoted to one decimal place. When the associated $1\sigma$ uncertainty from these 50 runs lies at or below the second decimal place (i.e., smaller than 0.05), only the median value is reported.}
    \label{tab:tension}
    \begin{tabularx}{\textwidth}{l
        >{\centering\arraybackslash}X
        >{\centering\arraybackslash}X
        >{\centering\arraybackslash}X
        >{\centering\arraybackslash}X
        >{\centering\arraybackslash}X
        >{\centering\arraybackslash}X}
        \toprule
        & \textbf{FABLE} 
        & \textbf{Long hardening time (5 Gyr)} 
        & \textbf{FABLE + macrophysical delays~\citepalias{buttigieg_2025}} 
        & \textbf{Boosted masses ($z_t = 1, \, f_b = 5$)} 
        & \textbf{Equal mass ratio (for $q > 0.01$)}  \\
        \midrule
        \rowcolor{gray!15}
        Power-law at $f = 1~\mathrm{yr}^{-1}$ & $2.2$ & 2.5 & $2.5$ & 2.0 & 1.9 \\
        Free spectrum (Bins: 2) & 1.6 & 2.5 & $2.1 \pm 0.1$ & 1.2 & 0.6 \\
        Free spectrum (Bins: 2, 3) & 1.7 & 2.5 & 2.4 & 1.2 & 0.5 \\
        Free spectrum (Bins: 2, 3, 4) & 1.1 & $2.2 \pm 0.1$ & $2.1\pm0.1$ & 0.7 & 0.2 \\
        Free spectrum (Bins: 2, 3, 4, 5) & 0.8 & $1.9\pm0.1$ & $1.7 \pm 0.1$ & $0.3$ & $0$ \\
        Free spectrum (Bins: 2, 3, 4, 5, 1) & 0.5 & $1.5\pm0.1$ & $1.2\pm 0.1$ & $0.2$ & 0 \\
        \bottomrule
    \end{tabularx}
\end{table*}

\section{Discussion \& Conclusions}
\label{sec:conclusion}

In this work, we used the population of merging SMBHBs from the \Fable{} cosmological simulation to predict the GWB detected by PTAs, employing the \texttt{holodeck} code to generate population-based strain spectra. Motivated by previous theoretical studies that somewhat underpredict the GWB amplitude with respect to that detected, we developed a quantitative method to measure the tension between predictions and PTA observations.

Applying this framework to the NANOGrav 15-year dataset, we find that the fiducial \Fable{} model slightly underpredicts the observed characteristic strain, though the discrepancy remains below $2.5\sigma$ for total binary hardening timescales shorter than $\sim 5$~Gyr. The hardening timescale itself remains one of the dominant uncertainties, as it depends sensitively on the stellar and gaseous environments of merging SMBHs. Rather than attempting to constrain it directly, we explore how physically plausible variations in the SMBHB population, such as enhanced SMBH growth at high redshift or preferential accretion onto secondary BHs, can offset the impact of longer, more conservative hardening timescales. These adjustments, all broadly consistent with current EM constraints and theoretical expectations, boost the predicted GWB amplitude and bring the \Fable{} model into close agreement with present PTA observations.

We find that the inferred level of tension depends on the data analysis techniques used to generate results from PTA observations. Since realistic SMBHB populations deviate from a $h_c(f) \propto f^{-2/3}$ power law, especially at high frequencies, tension estimates based on fixed-slope power-law fits can be misleading. Comparisons using the free-spectrum analysis, in which each frequency bin is treated independently, provide a more robust, model-agnostic measure of consistency. Note that although we do not explicitly calculate tension values using a power-law model with a free slope, this would yield tension values intermediate between the two approaches, reinforcing our conclusion that the \Fable{} predictions are statistically consistent with current observations.

We also assess the impact of finite cosmological volume on the predicted GWB using merger populations from the $(40 \, \mathrm{cMpc} \, h^{-1})^3$ and $(100 \, \mathrm{cMpc} \, h^{-1})^3$ \Fable{} boxes, as well as synthetic populations informed by the MXXL HMF to quantify realisation-to-realisation variance in these volumes. We find that the effect of limited volume on the GWB amplitude is frequency-dependent. At high frequencies, even the smaller \Fable{} box adequately captures the GWB signal. However, at lower frequencies, the smaller box underestimates the amplitude due to the absence of the most massive binaries which can make particularly large contributions to the GWB~\citep{raidal_2026}. In contrast, the larger \Fable{} box provides sufficient volume to statistically capture the amplitude at the most constraining frequency bin of PTA experiments ($f = 0.1 \, \mathrm{yr}^{-1}$), demonstrating convergence with respect to this observable. Furthermore, current PTA limits on individual sources disfavour the presence of extremely massive nearby binaries, reinforcing the conclusion that SMBHs within the mass range predicted by the simulations are the dominant contributors to the GWB.

We emphasise that the relatively low tension values found throughout this work reflect the effect of two important issues. Methodologically, free-spectrum analyses provide a more robust comparison to realistic SMBHB populations and generally yield lower tension values than power-law analyses. Astrophysically, \Fable{} predicts a relatively abundant population of high-mass BHs compared to EM-derived BHMFs. Because the GWB amplitude is dominated by the most massive binaries, this increases the predicted signal and contributes to the low tension values reported here.

In summary:
\begin{enumerate}
    \item \textbf{Quantifying tensions should be made with careful considerations.} The GWB signal from a realistic population of SMBHBs is expected to deviate from a fixed power-law, so comparisons using a power-law model can overestimate tensions with observations. Free-spectrum analyses provide a more robust basis for comparison, and we find that the tension inferred from the power-law analysis represents an upper bound relative to the free-spectrum results. Overall, the \Fable{} predictions show no significant tension with the NANOGrav 15-year data.
    \item \textbf{\Fable{} predicts a relatively high abundance of massive BHs.} The \Fable{} BHMF is in better agreement with BHMFs informed by the amplitude of the GWB, and is consistent with other state-of-the-art cosmological simulations. This contributes to lower tension values when compared to previous estimates made purely based on EM observations.
    \item \textbf{Astrophysical variations can reconcile models with data.} Modest and observationally consistent changes, such as enhanced secondary accretion or moderate boosts to SMBH masses at high redshift, could help amplify the predicted GWB amplitude, further decreasing the tension between observations and theory.
    \item \textbf{Limited cosmological volumes might impact the predicted GWB amplitude.} However, we find that a simulated box with a side length of $100 \, \mathrm{cMpc} \, h^{-1}$ is sufficient to capture the majority of SMBHs that dominate the GWB signal at the best-constrained frequencies.
\end{enumerate}

In the coming years, PTA datasets will continue to expand in both sensitivity and frequency coverage, complemented by increasingly detailed EM observations of the SMBH population. At the same time, theoretical models and cosmological simulations are becoming more sophisticated, but still carry significant astrophysical uncertainties. Recognising these uncertainties and ensuring that observational inferences are compared to theoretical predictions in a consistent and physically motivated way will be essential for building a coherent and comprehensive picture of SMBH evolution across cosmic time.

\section*{Acknowledgements}
We thank Giulia Ortame for helpful discussions on this work.
SB is supported by the Cambridge Centre for Doctoral Training in Data-Intensive Science, funded by the UK Science and Technology Facilities Council (STFC). DS acknowledges support from the STFC, grant code ST/W000997/1. MAB is supported by a UKRI Stephen Hawking Fellowship (EP/X04257X/1). This work performed using the following facilities: the Cambridge Service for Data Driven Discovery (CSD3) operated by the University of Cambridge Research Computing Service (www.csd3.cam.ac.uk), provided by Dell EMC and Intel using Tier2 funding from the Engineering and Physical Sciences Research Council (capital grant EP/P020259/1), and DiRAC funding from the STFC (www.dirac.ac.uk), as well as the DiRAC@Durham facility operated by the Institute for Computational Cosmology on behalf of the STFC DiRAC HPC Facility (www.dirac.ac.uk). 
The equipment was funded by BEIS capital funding via STFC capital grants ST/P002293/1, ST/R002371/1 and ST/S002502/1, Durham University, and STFC operations grant ST/R000832/1. DiRAC is part of the National e-Infrastructure.

\section*{Data Availability}

The \Fable{} data used in this paper is currently not publicly available through a web server, but is available upon request. The scripts used to run this analysis, together with the BH merger catalogues from \Fable{}, are available at \url{https://github.com/steph-buttigieg/gwbtensioncalc}.



\bibliographystyle{mnras}
\bibliography{references} 




\appendix

\section{Free spectrum analysis comparison}
\label{appendix:free_spectrum}
\begin{figure}
    \centering
    \includegraphics[width=\columnwidth]{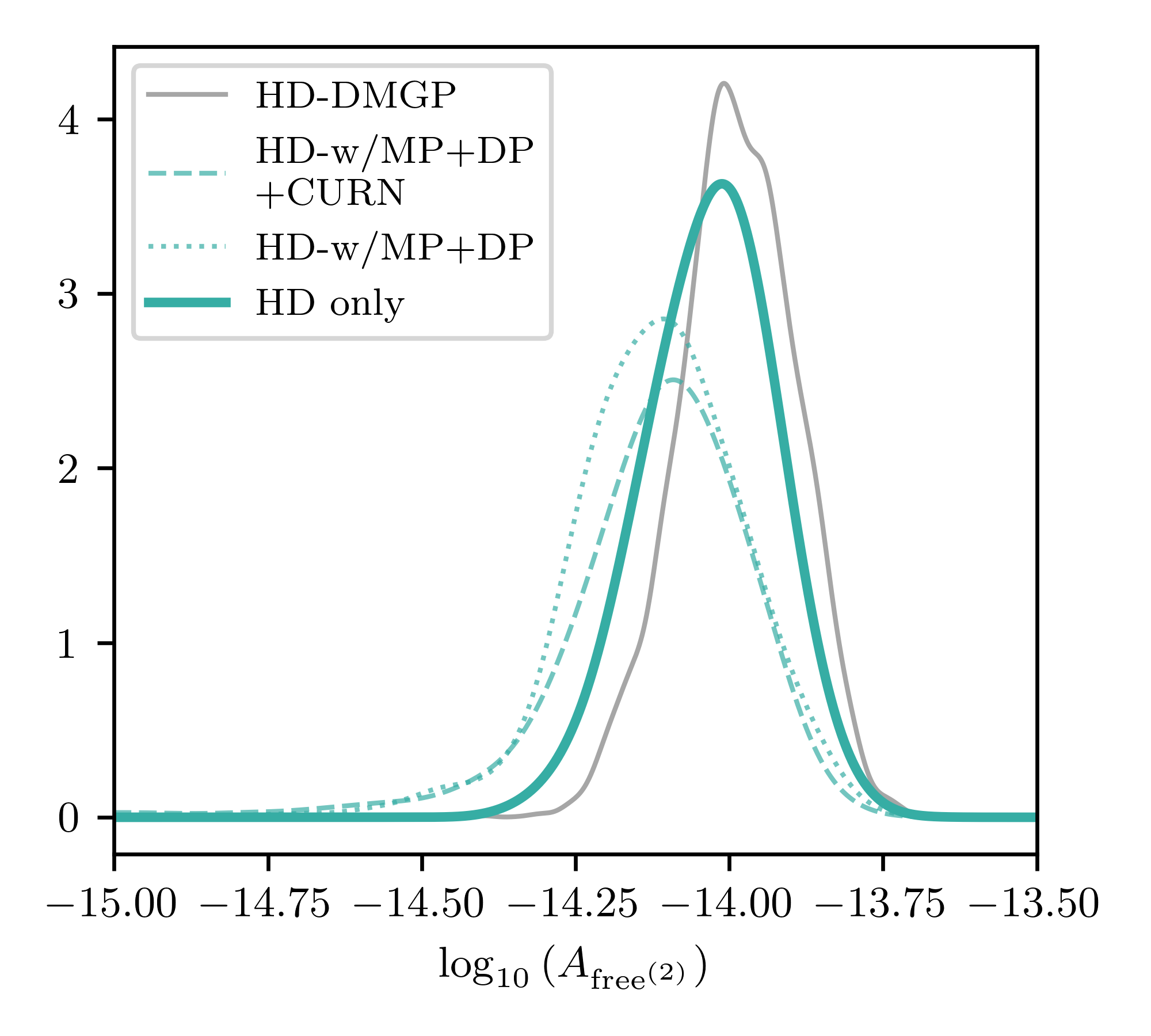}
    \caption{The posterior for the characteristic strain in the second lowest frequency bin of the free-spectrum analysis of the NANOGrav 15-year dataset. The teal solid line shows the results when assuming HD correlations only, the dashed teal line also fits for MP and DP (HD-w/MP+DP), while the dotted line also accounts for CURN (HD-w/MP+DP+CURN). The grey solid line fits for the HD correlation while also considering the dispersion measure (HD-DMGP).}
    \label{fig:free_spectrum}
\end{figure}

As discussed in Section~\ref{sec:tension}, several assumptions can be made when deriving posteriors in a free-spectrum analysis. In this work, we adopt the HD-correlated free spectrum, which represents the simplest model of a common red-noise process consistent with Hellings–Downs (HD) correlations. However, alternative analyses can include additional spatially correlated components such as the monopole (MP), dipole (DP), or common uncorrelated red noise (CURN) terms. These components typically reduce the inferred HD-correlated amplitude, since part of the low-frequency power may be absorbed by processes unrelated to the GWB. Another important consideration is variability in the ionised interstellar medium, represented by the dispersion measure (DM), which can introduce frequency-dependent timing noise~\citep{agazie_2023}. Fig.~\ref{fig:free_spectrum} compares posteriors for the characteristic strain in the second-lowest frequency bin under these different model assumptions from the NANOGrav 15-year dataset. Broader comparisons across all frequency bins can be found in Fig.~1 of \citet{agazie_2023}.

Using the difference vector method described in Appendix~\ref{appendix:difference_vector}, we quantify the tension between the HD-only posterior and the three alternative noise models in this frequency bin. We find a tension of $0.3\sigma$ between the HD and HD-DMGP posteriors, and $0.5\sigma$ (in the opposite direction) between the HD and HD-w/MP+DP posteriors. This indicates that, depending on the chosen noise model, the inferred tension with theoretical predictions in this frequency bin could increase by roughly $0.3\sigma$ or decrease by about $0.5\sigma$ in cases where the signal is underpredicted. Consequently, tension values presented in Section~\ref{sec:revisiting-tensions} should be interpreted with this caveat in mind.

\section{Building a tension calculator}
\label{appendix:difference_vector}

There exist several methods to quantify the tension between theoretical predictions of the GWB and the latest PTA measurements. As discussed in Section~\ref{sec:tension}, in this work we adopt the difference vector method introduced by \citet{battye_2015}. A detailed description of this method and its implementation is provided below.

Given two probability distributions, $P_1 (\vect{\theta_1})$ and $P_2(\vect{\theta_2})$, where $\vect{\theta_1}$ and $\vect{\theta_2}$ represent the parameters of each distribution, one can construct a new distribution for the parameter $\vect{\delta \theta} = \vect{\theta_1} - \vect{\theta_2}$ through the the following integral of the original distributions:
\begin{equation} \label{eqn:difference_vector}
P(\vect{\delta \theta}) = \int \mathrm{d}\vect{\theta'}P_1(\vect{\theta'})P_2(\vect{\theta'}-\vect{\delta \theta})\, .
\end{equation}
In practice, $P(\vect{\delta \theta})$ can be obtained by independently sampling from $P_1(\vect{\theta_1})$ and $P_2(\vect{\theta_2})$, and constructing the distribution of the differences between the sampled points.

\begin{figure*}
  \centering
  \begin{subfigure}[b]{\columnwidth}
    \centering
    \includegraphics[width=\columnwidth]{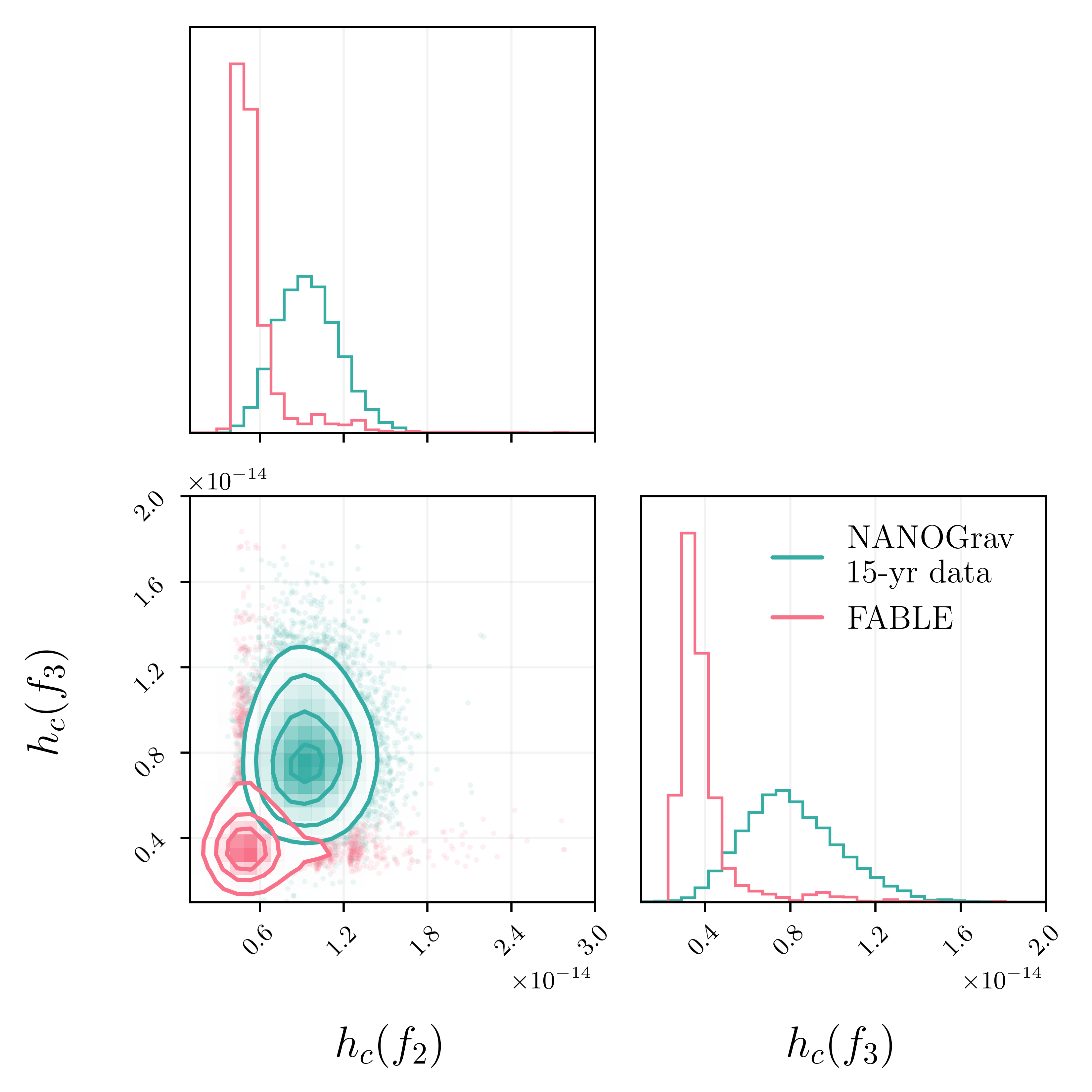}
    \label{fig:bhard_ranges}
  \end{subfigure}
  \hfill
  \begin{subfigure}[b]{\columnwidth}
    \centering
    \includegraphics[width=\columnwidth]{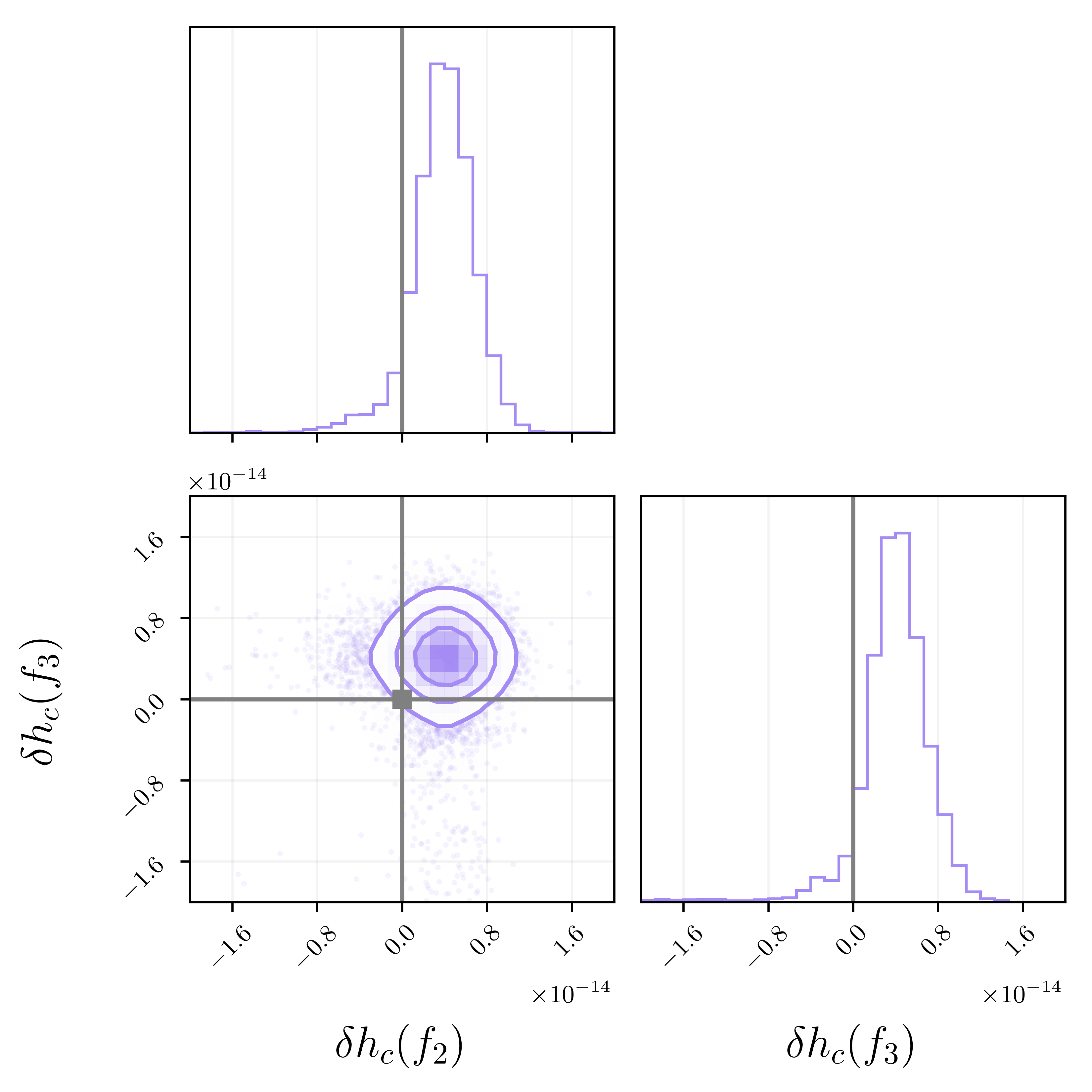}
    \label{fig:bhmd_ranges}
  \end{subfigure}
  \caption{Left: 2D distribution of the strain at the 2nd and 3rd lowest frequency bins, $h_c(f_2)$ and $h_c(f_3)$ respectively. The distribution in teal corresponds to the free-spectrum analysis posteriors of the NANOGrav 15-yr dataset in these frequency bins, while the pink distribution is generated from 1000 different realisations of the GWB generated from the discrete \Fable{} population using \texttt{holodeck}. Right: Distribution of the difference vector $\vect{\delta \theta}$. These distributions correspond to a tension value of $1.6\sigma$.}
  \label{fig:distribution}
\end{figure*}

Because the frequency bins in the NANOGrav free-spectrum analysis are uncorrelated, an $n$-dimensional distribution, $P_1(\vect{\theta_1})$ can be constructed by sampling independently from each frequency bin. Similarly, a second $n$-dimensional distribution can be obtained from the simulated GWB, $P_2(\vect{\theta_2})$, using \Fable{} and \texttt{holodeck} as described in Section~\ref{sec:methods_holodeck}. Fig.~\ref{fig:distribution} visualises these distributions using the 2nd and 3rd lowest frequency bins from the NANOGrav data (see also the violin plot in Fig.~\ref{fig:gwb_comparison}). The left-hand panel shows the original distributions $P_1(\vect{\theta_1})$ and $P_2(\vect{\theta_2})$, where $\vect{\theta_1} = \left(h_{c}^\mathrm{data}(f_2), h_c^\mathrm{data}(f_3)\right)$ and $\vect{\theta_2} = \left(h_{c}^\mathrm{theory}(f_2), h_c^\mathrm{theory}(f_3)\right)$. The right-hand panel shows the corresponding distribution $P(\vect{\delta \theta})$ from equation~\eqref{eqn:difference_vector}. If the two input distributions were identical, $P(\vect{\delta \theta})$ would peak at $(0,0)$, indicated by the grey lines in Fig.~\ref{fig:distribution}.

The tension between the two distributions is then evaluated by calculating the p-value at the origin, i.e.~the probability of observing a difference vector more extreme than $\vect{0}$ under $P(\vect{\delta \theta})$,

\begin{align}
    p = \int_{\{\vect{\delta\theta}|P(\vect{\delta\theta})<P(\vect{0})\}} \mathrm{d}(\vect{\delta\theta}) \; P (\vect{\delta \theta}) \, .
\end{align}

In higher dimensions, the p-value is most efficiently estimated via Monte Carlo (MC) sampling: by drawing samples from $P(\vect{\delta \theta})$ and counting the fraction that have lower probability density than $P(\vect{0})$. This p-value is converted to a z-score (or sigma value) assuming an equivalent one-dimensional Gaussian,
\begin{equation}
    \sigma = \Phi^{-1}\left(1-\frac{p}{2}\right) \, ,
\end{equation}
where $\Phi^{-1}$ is the inverse cumulative distribution function of the standard normal distribution. For the distributions shown in Fig.~\ref{fig:distribution}, this method yields a tension of $1.6\sigma$.

\begin{figure}
    \centering
    \includegraphics[width=\columnwidth]{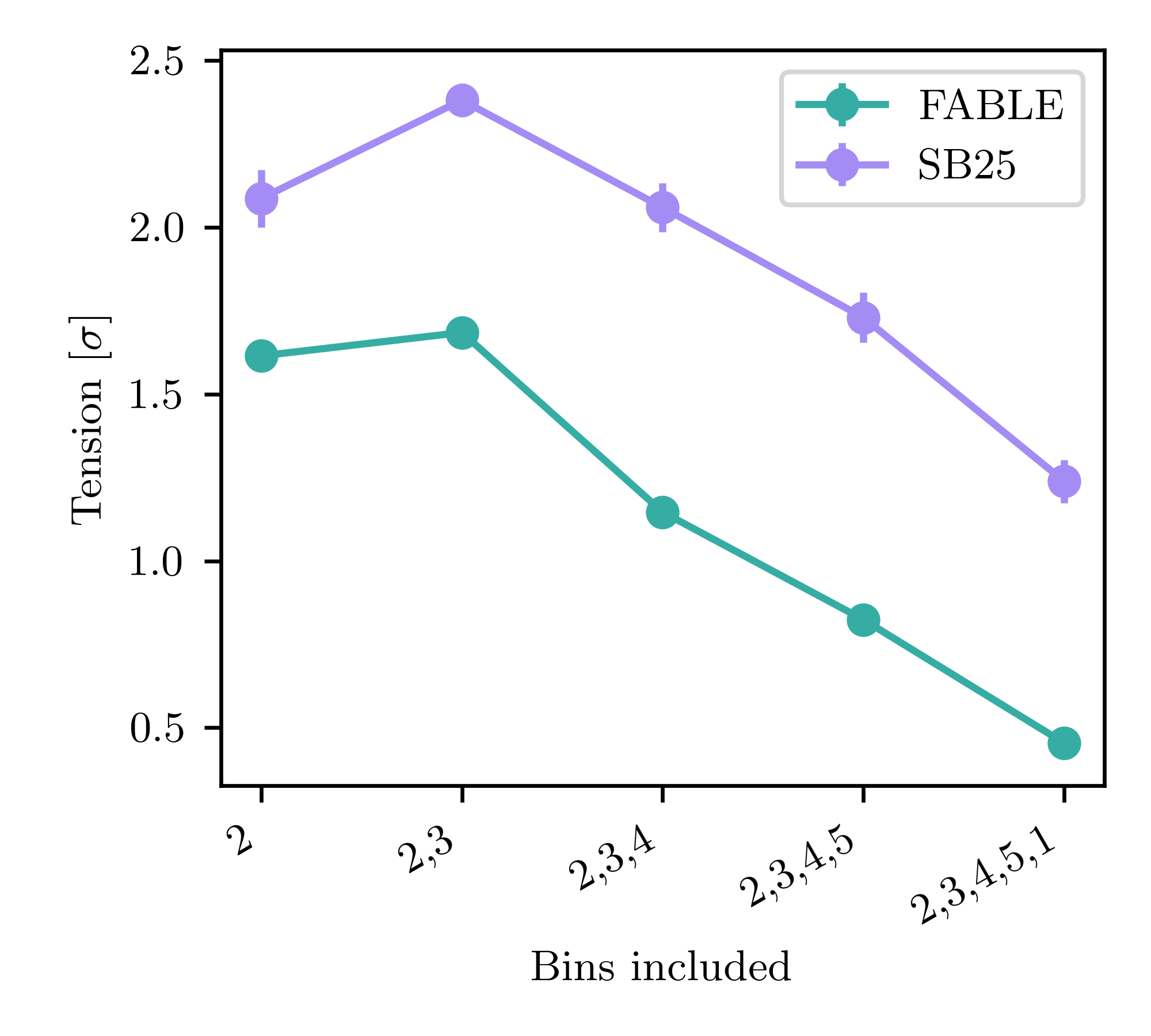}
    \caption{Tension values computed using the method described in Appendix~\ref{appendix:difference_vector}, for GWBs generated from the \Fable{} simulation (teal points) and following the approach in \citetalias{buttigieg_2025} with added macrophysical delays (purple points). These correspond to the GWBs shown in the same colours in the top-left panel of Fig.~\ref{fig:gwb_comparison}. The x-axis labels indicate which frequency bins are included in the tension calculation, with 2 denoting the second-lowest frequency bin in the free-spectrum analysis, and so on. The GWB distribution is built using 100,000 realisations, and tension values are calculated using $10^6$ Monte Carlo points and with 50 iterations. Solid dots show the median tension, and error bars represent the 68\% credible interval. These correspond to the values given in Table~\ref{tab:tension}.}
    \label{fig:tension}
\end{figure}

As evident from the comparison between the free-spectrum posteriors and the \Fable{} predictions in the top-left panel of Fig.~\ref{fig:gwb_comparison}, the resulting tension depends on which frequency bins are included in the analysis. To explore this, we apply the difference vector method to an increasing number of frequency bins, starting from the 2nd lowest, then successively adding the 3rd, 4th, 5th, and finally the 1st bin. The evolution of the resulting sigma values for both the unmodified \Fable{} simulation, and with the post-processing method introduced in \citetalias{buttigieg_2025} is shown in Fig.~\ref{fig:tension}.
For robustness, to calculate these tensions, we generate 100,000 independent realisations of the GWB to get a more representative statistical distribution from our various models. We then repeat the above procedure 50 times with a million Monte Carlo points and plot the median and $1\sigma$ confidence intervals of the resulting tensions. The highest tension occurs when only the 2nd and 3rd frequency bins are included; the significance decreases as less constraining bins are added, since these broaden the agreement between data and predictions. Consequently, the most conservative yet representative measure of the current data–model tension is obtained from the free-spectrum posteriors in the 2nd and 3rd lowest frequency bins, which we adopt as our fiducial calculation.

\section{Impact of mass boosting on the BHMF}\label{appendix:mass_boosting}
\begin{figure}
    \centering
    \includegraphics[width=\columnwidth]{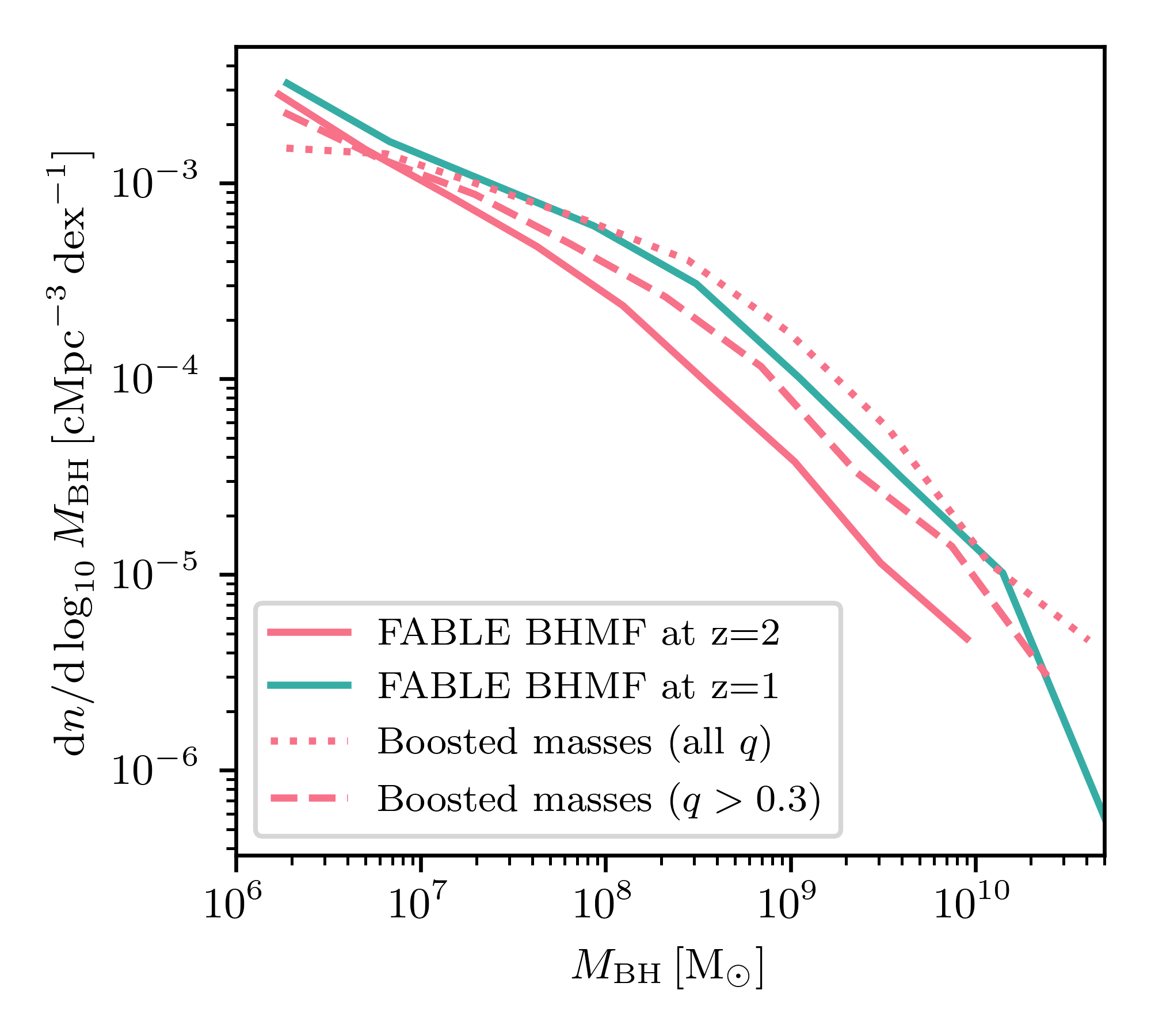}
    \caption{The BHMF at $z=1$ and $z=2$ directly from the \Fable{} simulation (teal and pink solid lines, respectively), showing the evolution of the BHMF in the simulation in this redshift interval. The dotted pink line shows the BHMF at $z=2$, modified by boosting the masses of BHs involved in any merger during this redshift interval. The dashed pink line shows the same result but considers only boosting BH masses for BHs involved in a merger with $q\geq0.3$. }
    \label{fig:bhmf_boost}
\end{figure}

In Section~\ref{sec:mass_boosting}, we apply a boost to the masses of BHs in early mergers to explore how enhanced SMBH growth might influence the predicted GWB. To avoid excessively altering the underlying BH mass function (BHMF), we restrict the boost to major mergers with mass ratio $q \geq 0.3$. Fig.~\ref{fig:bhmf_boost} illustrates the motivation for this choice.

In this figure, the solid lines show the BHMF in the \Fable{} simulation at $z=2$ (pink) and at $z=1$ (teal), representing the growth of the BH population over this redshift interval. We then estimate the fraction of BHs involved in mergers between $z=2$ and $z=1$ as a function of mass, boost the corresponding subset of BHs in the $z=2$ population by a factor of $f_b=5$, and reconstruct the resulting BHMF. The dotted line in Fig.~\ref{fig:bhmf_boost} shows the BHMF obtained when boosting all mergers regardless of $q$, while the dashed line shows the result when only major mergers ($q > 0.3$) are boosted. Comparing these to the original growth in the simulation (solid lines), we see that boosting all mergers overshoots the growth that happens in the simulation, whereas restricting the boost to major mergers results in a more conservative modification to the BHMF. It is worth emphasising here that this is a conservative estimate of the BHMF modification caused by the boosted high redshift BH masses: due to the self-regulated nature of BH feedback, these BHs with boosted masses would experience less subsequent growth at low redshifts, resulting in the local BHMF which does not depart much from FABLE predictions.   

Together with the results discussed in Section~\ref{sec:mass_boosting}, this shows that boosting a fraction of BHs at high redshift, which undergo major mergers, can enhance the contribution of these binaries to the GWB, increasing its amplitude without introducing growth that overshoots the evolution seen in the simulation itself. We also apply this procedure for a larger boosting factor, $f_b = 10$, and find that even when the boost is restricted to major mergers, the resulting BHMF evolution exceeds that seen in the \Fable{} simulation. This justifies our choice of $f_b = 5$ as our fiducial value as discussed in Sec.~\ref{sec:mass_boosting}.

\section{Semi-analytic model for building synthetic merger catalogues} \label{sec:appendix-convergence}
As discussed in Section~\ref{sec:volume}, the \Fable{} cosmological boxes might not be sufficiently large to capture BHs at the high-mass end of the BHMF, which might have a significant impact on the GWB. In this Appendix, we will describe the method used to extend the BHMF to higher masses informed by the HMF from the MXXL simulation, and how we then generate synthetic populations of SMBH mergers from these extended BHMFs. 

The MXXL simulation is a very large high-resolution N-body cosmological simulation that follows the non-linear growth of DM structures within a box size with side length of 4.11 cGpc (or $3 \, h^{-1} \, \mathrm{cGpc}$ with $h=0.73$)~\citep{angulo_2012}. At $z=0$, MXXL contains more than 700 million haloes with a mass larger than $1.7\times 10^{11}$~\Msun{}, making it incredibly useful for large-scale statistics. Since this large-volume simulation does not self-consistently model a population of SMBHs, we use the \Fable{} simulation to synthetically produce a population of SMBH mergers from the MXXL HMF.

By combining the results from the MXXL simulation with the smaller, higher resolution Millennium Simulation~\citep{springel_2005} and the Millennium-II Simulation~\citep{boylan_2009}, the HMF is analytically described by
\begin{equation} \label{eqn:hmf}
    M \frac{dn}{dM} = \rho_0 \frac{d \ln \sigma^{-1}}{dM}f(\sigma(M)),
\end{equation}
where $\rho_0$ is the mean mass density of the Universe and $\sigma(M)$ is the variance of the linear density field within a top-hat filter containing total mass $M$~\citep{angulo_2012}. The fitting function for subhaloes, i.e.~structures identified using the \textsc{subfind} algorithm, is given by
\begin{equation} \label{eqn:fitting_function}
    f(\sigma (M)) = 0.265 \times \left[ \frac{1.675}{\sigma(M)}+1 \right] ^{1.9} \exp \left[ \frac{-1.4}{\sigma ^2 (M)} \right] \,.
\end{equation}
Note that $\sigma (M)$ is a function of $z$, which introduces a redshift dependence to the HMF.

Our aim is to use the analytical fit of the MXXL HMF, which extends to higher masses than those directly sampled by the \Fable{} simulations, to extend the \Fable{} BHMF. We apply the method described below independently to both simulation volumes, \Fable{}-40 and \Fable{}-100, thereby constructing two versions of the extended BHMF that reflect the different statistics and mass ranges probed by each box.

For each simulation, we first calibrate the MXXL HMF to the corresponding \Fable{} HMF to account for differences in the underlying cosmology. This is achieved by multiplying equation~\eqref{eqn:hmf} by the mean ratio between the analytical fit and the \Fable{} HMF over the mass interval where the two overlap and where \Fable{} has sufficient statistics to robustly capture the underlying HMF. 

We then use \Fable{} to obtain a relation between BH mass and the total dark matter mass of the host subhalo. At a particular redshift, we fit the relation between host halo mass $M_{\mathrm{DM}}$ and BH mass $M_{\mathrm{BH}}$ with a broken power-law model, allowing us to capture the change in behaviour of the relation and, in particular, the shallower slope at the high-mass end. Only systems with $M_{\mathrm{BH}} \geq 10^{7}\,M_\odot$ are included to minimise numerical effects associated with the seeding prescription. The fit yields the parameters describing the low- and high-mass slopes and normalisation of the mean relation.

Finally, we use the calibrated BH–halo mass relation to transform the analytical HMF (equation~\eqref{eqn:hmf}) into an extended BHMF, thereby capturing the contribution of higher-mass systems not directly resolved in \Fable{}. The resulting extended BHMFs, together with those measured directly in \Fable{}-40 and \Fable{}-100, at redshift zero are shown in Fig.~\ref{fig:bhmf}.

To construct synthetic BH merger catalogues from the extended BHMF, we first bin all \Fable{} merger events into redshift intervals. These bins are adaptive and are chosen to balance the number of merger events they contain with the availability of snapshot data, which differs between the two simulation volumes. A synthetic population is then generated through the following procedure:

\begin{enumerate}
    \item \textbf{Sampling the BH population.} For each redshift interval, we sample a population of BHs from the extended BHMF representing the BH population present at the beginning of the interval for the specified synthetic volume. This sampled population constitutes one realisation of the BH population expected within that volume, given the underlying BHMF.

    \item \textbf{Estimating the merger incidence.} In bins of primary BH mass, we compute the ratio between the number of mergers occurring within the redshift interval and the number of BHs present at its start in the \Fable{} simulation. This provides an estimate of the expected number of mergers per BH as a function of primary mass. We fit a power law to these measurements and use it to extrapolate the merger incidence to higher masses that are not directly sampled in the simulations. We find that the expected number of mergers per BH increases with primary mass. This trend reflects the hierarchical growth of structure, whereby more massive BHs reside in more massive haloes that experience higher merger rates and host richer satellite populations, leading to an enhanced probability of merger events.

    \item \textbf{Modelling the mass ratio distribution.} Using the same primary mass bins, we characterise the distribution of mass ratios as a function of primary mass. In bins with sufficient merger statistics, we construct KDEs of the mass ratio distribution. We also fit the mean trend with primary mass and extrapolate it beyond the sampled range to describe mergers involving more massive systems. We find that the mean mass ratio decreases with increasing primary mass, indicating that more massive black holes preferentially grow through mergers with lower-mass companions.

    \item \textbf{Modelling the redshift distribution.} We construct a KDE describing the distribution of merger redshifts within each redshift interval.

    \item \textbf{Sampling the number of mergers.} For each BH sampled from the BHMF, we compute the expected number of mergers from the fitted merger incidence relation and draw the realised number of mergers from a Poisson distribution with this mean.

    \item \textbf{Building merger histories.} For each BH, we construct a merger history by sequentially sampling merger events. Mass ratios are drawn from the KDEs when the BH mass lies within the range constrained by the simulations; otherwise, they are drawn from a log-normal distribution whose mean follows the extrapolated trend and whose variance is set by the average variance measured in the well-sampled bins. After each merger, the primary BH mass is updated accordingly.

    \item \textbf{Assigning initial separations.} We assign an initial binary separation by sampling from KDEs constructed from the simulation data, using the highest-mass bin when the system lies outside the sampled range.
\end{enumerate}

This procedure can be applied to any synthetic simulation volume by sampling the BH population directly from the BHMF at the corresponding volume, keeping in mind that merger properties of BHs at the high mass end of the BHMF are extrapolated from the lower-mass population in \Fable{}. Populations generated in this way represent different realisations of the underlying BH distribution and therefore allow us to probe the impact of cosmic variance, under the assumption that the merger statistics are unchanged. Results from this method are discussed in Section~\ref{sec:volume}.


\bsp	
\label{lastpage}
\end{document}